\documentclass[aps,prd,twocolumn,preprintnumbers,
superscriptaddress,nofootinbib]{revtex4-2}
\usepackage{dcolumn}
\usepackage{amsmath}
\usepackage{amssymb}
\usepackage{amsthm}
\usepackage{mathtools}
\usepackage{mathrsfs}
\usepackage{bm}
\usepackage{slashed} 
\usepackage{graphicx}
\usepackage{multirow}
\usepackage{tikz}

\usepackage{natbib}
\usepackage[caption=false]{subfig}
\usepackage{relsize}	
\usepackage{array}
\usepackage{float}
\usepackage{color}
\usepackage{xcolor}
\usepackage{soul}
\usepackage{verbatim} 
\usepackage{siunitx}
\newcommand{\ket}[1]{|\,#1\,\rangle}          

\newcommand{\LCp}{{\scriptscriptstyle +}}

\usepackage{hyperref}
\usepackage{CJKfntef}
\usepackage{ulem}
\hypersetup{colorlinks=true,linkcolor=purple,anchorcolor=blue,citecolor=blue, filecolor=blue,urlcolor=red,bookmarksnumbered=true,
	pdfview=FitB
}
\allowdisplaybreaks
\def\be{\begin{eqnarray}}
	\def\ee{\end{eqnarray}}

\colorlet{purple1}{blue!70!red}
\colorlet{darkred}{red!50!black}
\usepackage{appendix}

\begin{document}

\title{Basis light-front quantization approach to $\Lambda$ and $\Lambda_c$ and their isospin triplet baryons}

\author{Tiancai~Peng}
\email{pengtc20@lzu.edu.cn} 
\affiliation{School of Physical Science and Technology, Lanzhou University, Lanzhou 730000, China}
\affiliation{Research Center for Hadron and CSR Physics, Lanzhou University and Institute of Modern Physics of CAS, Lanzhou 730000, China}

\author{Zhimin~Zhu}
\email{zhuzhimin@impcas.ac.cn} 
\affiliation{Institute of Modern Physics, Chinese Academy of Sciences, Lanzhou 730000, China}
\affiliation{School of Nuclear Science and Technology, University of Chinese Academy of Sciences, Beijing 100049, China}

\author{Siqi~Xu}
\email{xsq234@impcas.ac.cn} 
\affiliation{Institute of Modern Physics, Chinese Academy of Sciences, Lanzhou 730000, China}
\affiliation{School of Nuclear Science and Technology, University of Chinese Academy of Sciences, Beijing 100049, China}

\author{Xiang Liu}\email{xiangliu@lzu.edu.cn}
\affiliation{School of Physical Science and Technology, Lanzhou University, Lanzhou 730000, China}
\affiliation{Research Center for Hadron and CSR Physics, Lanzhou University and Institute of Modern Physics of CAS, Lanzhou 730000, China}
\affiliation{Lanzhou Center for Theoretical Physics, Key Laboratory of Theoretical Physics of Gansu Province and Frontier Science Center for Rare Isotopes, Lanzhou University, Lanzhou 730000, China}

\author{Chandan~Mondal}
\email{mondal@impcas.ac.cn} 
\affiliation{Institute of Modern Physics, Chinese Academy of Sciences, Lanzhou 730000, China}
\affiliation{School of Nuclear Science and Technology, University of Chinese Academy of Sciences, Beijing 100049, China}

\author{Xingbo~Zhao}
\email{xbzhao@impcas.ac.cn} 
\affiliation{Institute of Modern Physics, Chinese Academy of Sciences, Lanzhou 730000, China}
\affiliation{School of Nuclear Science and Technology, University of Chinese Academy of Sciences, Beijing 100049, China}

\author{James~P.~Vary}\email{jvary@iastate.edu}
\affiliation{Department of Physics and Astronomy, Iowa State University,
	Ames, IA 50011, U.S.A.}

\collaboration{BLFQ Collaboration}
\date{\today}

\begin{abstract}
We obtain the masses, the electromagnetic properties, and the parton distribution functions (PDFs) of $\Lambda$, $\Lambda_c$, and their isospin triplet baryons, i.e, $\Sigma^0$, $\Sigma^+$, $\Sigma^-$ and $\Sigma_c^0$, $\Sigma_c^+$, $\Sigma_c^{++}$ from a light-front effective Hamiltonian in the leading Fock sector in the basis light-front quantization framework. The light-front wave functions of these baryons are given by the eigenstates of the effective Hamiltonian consisting of a three-dimensional confinement potential and a one-gluon exchange interaction with fixed
coupling. The masses of these baryons in our approach are in the experimental range while isospin-dependent mass differences are too small. Meanwhile, the electromagnetic properties are in agreement with the available experimental data, the lattice QCD simulations, and the other theoretical calculations. We also present the gluon and the sea quark PDFs, which we generate dynamically from the QCD evolution of the valence quark distributions.
\end{abstract}

\maketitle
\section{Introduction} \label{intro}
Although quantum chromodynamics (QCD) is the well-established theory for the strong interactions~\cite{Callan:1977gz}, where hadrons are built up from quarks and gluons, due to our incomplete understanding of the color confinement it is not yet possible to forecast the experimentally observed hadron spectroscopy from QCD first principles. Meanwhile, a successful theoretical framework for achieving valuable insights into hadron spectra and revealing partonic structures is provided by the  Hamiltonian formulation of QCD quantized on the light front~\cite{Brodsky:1997de,Bakker:2013cea}. Complementary insights into nonperturbative QCD can be accomplished by light-front holography~\cite{Brodsky:2014yha,Brodsky:2006uqa,deTeramond:2005su,deTeramond:2008ht,Ahmady:2021yzh,Ahmady:2021lsh}. For a practical approach, basis light-front quantization (BLFQ), which is based on the Hamiltonian formalism, provides a computational framework to solve relativistic many-body bound state problems in quantum field theories~\cite{Vary:2009gt,Zhao:2014xaa,Wiecki:2014ola,Li:2015zda,Jia:2018ary,Qian:2020utg,Tang:2019gvn,Mondal:2019jdg,Xu:2021wwj,Liu:2022fvl,Nair:2022evk,Lan:2021wok,Adhikari:2021jrh,Mondal:2021czk,Hu:2020arv,Lan:2019img,Lan:2019rba,Lan:2019vui}.

Electromagnetic form factors (EMFFs) and parton distribution functions (PDFs) are two essential probes of the internal structure of bound states. Both observables deepen our understanding of nonperturbative and perturbative QCD effects encoded in hadrons. The Fourier transform of the EMFFs provides information about spatial distributions such as the charge and the magnetization distributions inside the hadron, whereas the PDFs encode the nonperturbative structure of the hadron in terms of the distribution of longitudinal momentum and polarization carried by the quarks and gluons as its constituents.
While the nucleon's EMFFs and PDFs attract numerous dedicated experimental and theoretical efforts for several decades and are becoming better known (see Ref.~\cite{Xu:2021wwj} and  references therein), our information about the partonic structures of the $\Lambda$, $\Lambda_c$, and their isospin triplet baryons is very limited. This is due to the short lifetimes of these baryons making them unfeasible as targets. Only the magnetic moments of some of these baryons have been experimentally determined. Meanwhile, the timelike EMFFs of the $\Lambda$ baryon have been measured with good precision~\cite{DM2:1990tut,BaBar:2007fsu,BESIII:2017hyw}. The EMFFs of the $\Lambda$ hyperons have been analyzed by several theoretical studies~\cite{Kubis:2000aa,Puglia:2000jy,Julia-Diaz:2004yqv,Lin:2008mr,VanCauteren:2003hn,Haidenbauer:2016won,Yang:2017hao,Dalkarov:2009yf,Faldt:2017kgy,Faldt:2016qee,Baldini:2007qg}.

The first deep inelastic scattering (DIS) experiment at SLAC~\cite{Bloom:1969kc} showed the partonic substructure of the nucleon. From DIS processes~\cite{Bjorken:1969ja}, one can extract the PDFs, which encode the nonpertubative structure of the hadrons in terms of the number densities of their confined constituents.  The PDFs are functions of the light-front longitudinal momentum fraction ($x$) of the hadron carried by the
constituents. At the leading twist, the complete spin structure of the spin-$1/2$ hadrons is described in terms of three independent PDFs, namely, the unpolarized $f_1(x)$, the helicity $g_1(x)$, and the transversity $h_1(x)$. The global fitting collaborations such as HERAPDF \cite{Alekhin:2017kpj}, NNPDF \cite{NNPDF:2017mvq}, MMHT \cite{Harland-Lang:2014zoa}, CTEQ \cite{Dulat:2015mca}, and MSTW \cite{Martin:2009iq} have made considerable efforts to determine nucleon PDFs and their uncertainties. Meanwhile, the nucleon PDFs have also been investigated using different theoretical approaches~(see Ref.~\cite{Xu:2021wwj} and  references therein). On the other hand, much less information is available on the PDFs of the $\Lambda$, $\Lambda_c$, and their isospin triplet baryons~\cite{Gockeler:2002uh,Boros:1999da,Ellis:2002zv,Ma:1999wp,Ma:1999gj,Anselmino:2001js,Santopinto:2010zza}, while precise knowledge of PDFs is required for the analysis and interpretation of the scattering experiments in the LHC era. 

In this paper, with the theoretical framework of BLFQ~\cite{Vary:2009gt}, we adopt an effective light-front Hamiltonian~\cite{Mondal:2019jdg,Xu:2021wwj} and solve for the resulting mass eigenstates for the $\Lambda$, $\Lambda_c$, and their isospin triplet baryons at
the scales suitable for low-resolution probes. With quarks as the only explicit degrees of freedom, our effective Hamiltonian incorporates a three-dimensional (transverse and longitudinal) confinement potential and the one-gluon exchange (OGE) interaction that account for the dynamical spin effects~\cite{Li:2015zda}. By solving this Hamiltonian in the leading Fock space, using the quark masses, the strength of confinement, and the coupling constant as fitting parameters, we determine the masses of the baryons as the eigenvalues of the Hamiltonian. We also obtain the desired light-front wave functions (LFWFs) of the baryons as the eigenfunctions of the Hamiltonian. We then employ the LFWFs to study the electromagnetic properties and the PDFs of those baryons. We compare our BLFQ computations for the EMFFs, magnetic moments, and charge radii of the $\Lambda\,(\Sigma^0,\,\Sigma^+,\,\Sigma^-)$ with available experiments and with other theoretical approaches~\cite{Kubis:2000aa,Puglia:2000jy,Julia-Diaz:2004yqv,Lin:2008mr,VanCauteren:2003hn}. The experimental data are not yet available for the electromagnetic properties of the  $\Lambda_c\,(\Sigma_c^+,\,\Sigma_c^{++},\,\Sigma_c^0$). We compare our results of the magnetic moments and the charge radii of these baryons with other theoretical calculations reported in Refs.~\cite{Julia-Diaz:2004yqv,Faessler:2006ft,Sharma:2010vv,Barik:1983ics,Bernotas:2012nz,Zhu:1997as,Kumar:2005ei, Patel:2007gx,Yang:2018uoj,Wang:2018gpl}. 

This paper is organized as follows. A brief description
of the BLFQ formalism and the light-front effective
Hamiltonian for the baryons is discussed in Sec.~\ref{sec:hami}. We discuss the electromagnetic properties of the $\Lambda$, $\Lambda_c$, and their isospin triplet baryons in Sec.~\ref{sec:EMFFs}, whereas their PDFs are presented in Sec.~\ref{sec:pdfs}. We provide a brief summary
and conclusion in Sec.~\ref{sec:conclusion}.

\section{BLFQ framework for the baryons}\label{sec:hami}
The central task of the BLFQ approach is to solve the following eigenvalue equation to obtain the mass spectrum and the LFWFs of hadronic bound states
\begin{align}
P^2\vert \Psi\rangle=M^2\vert\Psi\rangle\,,\label{eq:LF_Schrodinger}
\end{align}
where $P^2=P^+P^--P_\perp^2$ is the effective light-front Hamiltonian and the operators $P^- $, $P^+$ and $P_{\perp}$ are the light-front quantized Hamiltonian, longitudinal momentum, and the transverse momentum, respectively of the system. Using a suitable matrix representation for the Hamiltonian, the diagonalization of Eq.~(\ref{eq:LF_Schrodinger}) generates the eigenvalues $M^2$, which correspond to the mass squared spectrum, and the associated eigenstates $\vert\Psi\rangle$ that encode structural information of the bound states.

In this paper, we solve the baryonic bound-state problem in the BLFQ framework using an effective light-front Hamiltonian defined below. At fixed light-front time, the baryonic state can be expressed in terms of various quark ($q$), antiquark ($\bar q$) and gluon $(g)$ Fock components,
\begin{align}\label{Eq1}
|B\rangle=&\psi_{(3q)}|qqq\rangle+\psi_{(3q+q\bar q)}|qqqq\bar q\rangle +\psi_{(3q+1g)}|qqqg\rangle+\dots\, , 
\end{align}
where the $\psi_{(\dots)}$ represent the probability amplitudes to obtain the different parton configurations in the baryon. Within BLFQ, each Fock sector itself consists of an
infinite number of basis states. For the purpose of numerical simulations, we employ both a Fock-sector truncation
and limits on the basis states within each Fock sector.
Here, we restrict ourselves only to the valence Fock component to describe the valence quark contribution to baryon properties.

We adopt the effective Hamiltonian ($H_{\mathrm{eff}}=P^+P^-$) that incorporates the holographic QCD confinement potential supplemented by the longitudinal confinement, and the OGE interactions~\cite{Mondal:2019jdg,Xu:2021wwj}
\begin{align}\label{hamilton}
H_{\rm eff}=&\sum_i \frac{{\vec p}_{\perp i}^2+m_{i}^2}{x_i}+\frac{1}{2}\sum_{i\ne j}\kappa^4 \Big[x_ix_j({ \vec r}_{\perp i}-{ \vec r}_{\perp j})^2\nonumber\\
&-\frac{\partial_{x_i}(x_i x_j\partial_{x_j})}{(m_{i}+m_{j})^2}\Big]
+\frac{1}{2}\sum_{i\ne j} \frac{C_F 4\pi \alpha_s}{Q^2_{ij}}\nonumber\\&\times \bar{u}_{s'_i}(k'_i)\gamma^\mu{u}_{s_i}(k_i)\bar{u}_{s'_j}(k'_j)\gamma^\nu{u}_{s_j}(k_j)g_{\mu\nu},
\end{align}
where $x_i$ and ${\vec p}_{\perp i}$ represent the longitudinal momentum fraction and the relative transverse momentum carried by quark $i$. $m_{i}$ is the mass of the quark $i$, and $\kappa$ determines the strength of the confinement. The variable $\vec{r}_\perp={ \vec r}_{\perp i}-{ \vec r}_{\perp j}$ defines the transverse separation between two quarks. The last term in the effective Hamiltonian represents the OGE interaction with $Q^2_{ij}=-q^2=-(1/2)(k'_i-k_i)^2-(1/2)(k'_j-k_j)^2$ being the average momentum transfer squared, $C_F =-2/3$ corresponds to the color
factor, $\alpha_s$ defines the coupling constant, and $g_{\mu\nu}$ is the metric tensor. ${u}(k_i,s_i)$ corresponds to the spinor with momentum $k_i$ and spin $s_i$ of the parton in the baryon.  We have neglected electromagnetic interactions among the quarks.

We have assumed the same strength for our transverse and longitudinal confinement based on the fact that this form reduces to a symmetric 3-dimensional harmonic potential in the non-relativistic limit~\cite{Li:2015zda}.  It is possible that relaxing this assumption to allow for independent longitudinal and transverse confinement strengths (which therefore introduces another phenomenological parameter) could be advantageous~\cite{Li:2021jqb, Li:2022mlg}. We will pursue this additional freedom in a future work.

The basis states of each Fock particle are represented in
terms of the transverse and longitudinal coordinates along
with the helicity quantum numbers~\cite{Zhao:2014xaa}. We exclude the color degree of freedom in the current formalism since, for the pure valence sector considered as a color singlet, a color factor suffices when combined with the strength of our effective
OGE interaction.
Following BLFQ~\cite{Vary:2009gt,Zhao:2014xaa}, we expand $\ket{\Psi}$ in terms of the two dimensional harmonic oscillator (2D-HO) basis state in the transverse direction and the discretized plane-wave basis in the longitudinal direction.

The longitudinal momentum of the particle is characterized by the quantum number $k$. The longitudinal coordinate $x^-$ is confined to a box of length $2L$ with anti-periodic boundary conditions for fermions. The single-quark LFWF in the longitudinal coordinate space is then given by
\begin{align}\label{Kmax}
	\psi_{k}(x^-)=\frac{1}{2L}e^{i\frac{\pi}{L}kx^-}.
\end{align}
and the longitudinal momentum $p^\LCp=2\pi k/L$ is discretized, where the dimensionless quantity $k=\frac{1}{2}, \frac{3}{2}, \frac{5}{2}, ...$
All many-body basis states are selected to have the fixed total longitudinal momentum $P^+=\sum_ip_i^+$, where the sum is over the three quarks. We rescale $P^+$ using $K=\sum_i k_i$ such that $P^+=\frac{2\pi}{L}K$. For a given quark $i$, the longitudinal momentum fraction $x$ is then defined as $x_i=p_i^+/P^+=k_i/K.$

In the transverse direction, we employ the 2D-HO basis state, $\phi_{nm}(\vec{p}_\perp;b)$, which is characterized by two quantum numbers $n$ and $m$ representing the radial excitation and the angular momentum projection, respectively, of the Fock particle. In momentum space, the orthonormalized 2D-HO wave functions read~\cite{Vary:2009gt,Zhao:2014xaa}
\begin{align}
\phi_{n,m}(\vec{p}_{\perp};b)
 =&\frac{\sqrt{2}}{b(2\pi)^{\frac{3}{2}}}\sqrt{\frac{n!}{(n+|m|)!}}e^{-\vec{p}_{\perp}^2/(2b^2)}\nonumber\\
  &\times\left(\frac{|\vec{p}_{\perp}|}{b}\right)^{|m|}L^{|m|}_{n}(\frac{\vec{p}_{\perp}^2}{b^2})e^{im\theta},\label{ho}
\end{align}
with $b$ being its scale parameter with the dimension of mass; $L^{\alpha}_{n}(x)$ represent the generalized  Laguerre polynomials and $\theta={\rm arg}(\vec{p}_{\perp}/b)$. For the spin degrees of freedom, the quantum number $\lambda$ is used to define the helicity of the particle. Thus, each single-particle basis state is associated with four quantum numbers, $\{x,n,m,\lambda\}$. In addition, we have well defined values of the total angular momentum projection
$
M_J=\sum_i\left(m_i+\lambda_i\right)
$
for our multi-body basis states.

Beyond the Fock space truncation, within each Fock component, further truncation is still required to reduce the basis to a finite dimension. We reduce the infinite basis by introducing a truncation parameter $K$ in the longitudinal direction, and, in the transverse direction, we retain states with the total transverse quantum number 
\begin{align}
N_\alpha=\sum_i (2n_i+| m_i |+1),
\end{align}
satisfying $N_\alpha \le N_{\text{max}}$, where $N_{\text{max}}$ is the truncation parameter. The $N_{\rm{max}}$ controls the transverse momentum covered by the 2D-HO basis functions whereas $K$ is the basis resolution in the longitudinal direction. The $N_{\rm max}$ truncation naturally provides ultraviolet (UV) and infrared (IR) regulations. In momentum space, the UV regulator, $\Lambda_{\rm UV} \simeq b\sqrt{N_{\rm max}}$, while the IR regulator, $\lambda_{\rm IR} \simeq b/\sqrt{N_{\rm max}}$. The UV (IR) regulator increases (decreases)  with increasing the $N_{\text{max}}$~\cite{Zhao:2014xaa}, and both UV and IR increase as the HO basis scale parameter $b$ increases. 

We set up our basis using single-particle coordinates. The advantage of using these coordinates is that we can treat each particle in the Fock space on an equal footing~\cite{Zhao:2014xaa}. Meanwhile, $H_{\rm eff}$ includes the transverse center-of-mass (c.m.) motion, which is mixed with intrinsic motion. We introduce a constraint term 
\begin{align}
    \label{Hprime}
    H^{\prime}=\lambda_{L} ( H_{\rm c.m.}-2b^2 I)\, ,
\end{align}
into the effective Hamiltonian in order to factorize out the transverse c.m. motion from the intrinsic motion. We subtract the zero-point energy $ 2b^2 $ and multiply a Lagrange multiplier $\lambda_{L}$. $I$ denotes the identity operator. The c.m. motion is controlled by~\cite{Wiecki:2014ola},
\begin{align}
    H_{\rm c.m.}=\left(\sum_i \vec{p}_{i\perp}\right)^2+b^4\left(\sum_i x_i  \vec{r}_{i\perp}\right)^2 . 
\end{align}
When $\lambda_L$ is sufficiently large and positive, we are able to move the excited states of c.m. motion to higher energy than the low-lying spectrum of interest. Thus, the effective Hamiltonian we diagonalize is
\begin{align}
  H_{\rm eff}^{\prime} = H_{\rm eff} - \left(\sum_i \vec{p}_{i\perp}\right)^2 + \lambda_L (H_{\rm c.m.} - 2b^2I).
\end{align}

Upon diagonalization of this effective Hamiltonian matrix $H_{\rm eff}^{\prime}$ within the BLFQ bases, we obtain the eigenvalues that represent the mass spectrum. We also obtain the eigenvectors that correspond to the LFWFs in the BLFQ basis and provide the structural information of the systems. 
%
The resulting valence LFWF in momentum space is then expressed as an expansion in the orthonormal basis set consistent with the symmetries of the effective Hamiltonian
\begin{align}
\Psi^{\Lambda}_{\{x_i,\vec{p}_{i\perp},\lambda_i\}}&=\langle P, {\Lambda}|\{x_i,\vec{p}_{i\perp},\lambda_i\}\rangle\nonumber \\&=\sum_{\{n_i,m_i\}}\big( \psi^{\Lambda}_{\{x_{i},n_{i},m_{i},\lambda_i\}} \prod_i \phi_{n_i,m_i}(\vec{p}_{i\perp};b) \big),\label{wavefunctions}
\end{align}
with $\psi^{\Lambda}_{\{x_{i},n_{i},m_{i},\lambda_i\}}=\langle {P, {\Lambda}|\{x_i,n_i,m_i,\lambda_i\}}\rangle$ as the LFWF in BLFQ, where $P$ and $\Lambda$ indicate the momentum and the light-front helicity of the system. 
$\Lambda$and $\Sigma$ (or $\Lambda_c$and $\Sigma_c^+$) are represented as the ground and the first excited states of the effective Hamiltonian and thus the corresponding eigenvectors appearing in the LFWF Eq.~(\ref{wavefunctions}). And we can identify $\Lambda$ and $\Sigma$ (or $\Lambda_c$and $\Sigma_c^+$ ) by the flavour$\otimes$spin wavefunction in the appendix A.

 \begin{table*}[ht]
	\caption{Model parameters for the basis truncations $N_{\rm{max}}=8$ and $K=16.5$ for $\Lambda$ and $\Lambda_c$ baryons.}\label{tab:parameter}
	\centering 
		\begin{tabular}{|c|ccccc|}
			\hline \hline
			~~&~~  $\alpha_s$ ~~&~~ $m_{q/k}/m_{q/g}$ ~~&~~  $m_{s/k}/m_{s/g}$ ~~&~~$m_{c/k}/m_{c/g}$~~&~~ $\kappa$  \\ 
			\hline 		
			$\Lambda$ ~~&~~  1.06 $\pm0.1$ ~~&~~ 0.30/0.20~[GeV] ~~&~~ 0.39/0.29~[GeV] ~~&~~ -- ~~&~~  0.337~[GeV] \\ 
			\hline 
			$\Lambda_c$ ~~&~~  0.57$\pm0.06$ ~~&~~  0.30/0.20~[GeV] ~~&~~ -- ~~&~~ 1.58/1.48~[GeV] ~~&~~ 0.337~[GeV] \\ 
			\hline \hline 	
		\end{tabular}
\end{table*}

\section{Mass spectra}
There are four parameters in our model: the quark mass in the kinetic energy ($m_{q/k}$), the quark mass in the OGE interaction ($m_{q/g}$), the strength of confining potential ($\kappa=\kappa_T=\kappa_L$), and the coupling constant ($\alpha_s$) in the OGE interaction~\cite{Mondal:2019jdg,Xu:2021wwj}.  
We now outline our reasoning for flexibility in the choice of the vertex mass. In particular, our model features an effective OGE interaction that reflects short distance physics. It approximately describes the processes where valence quarks absorb and emit a gluon during which the system fluctuates between the $\ket{qqq}$, $\ket{qqqg}$, and higher Fock components. According to the mass evolution in renormalization group theory, the dynamical OGE would also produce contributions to the quark mass emerging from higher momentum scales leading to a decrease in the quark mass from the gluon dynamics. In turn, this leads to the suggestion that the quark mass in the OGE interaction would be lighter than the kinetic mass. The latter is associated with the long-range physics in our effective Hamiltonian. A similar treatment is also adopted in the literature~\cite{Brisudova:1994it,Burkardt:1998dd,Burkardt:1991tj}. 

We select the truncation parameters $N_{\rm{max}}=8$ and $K=16.5$, and the model parameters are summarized in Table~\ref{tab:parameter}. Note that the light quark mass ($m_q$) and the strength of confining potential were fixed by fitting the nucleon mass and the flavor form factors (FFs)~\cite{Mondal:2019jdg,Xu:2021wwj}. In this work, we replace one of the light quark masses by the effective strange (charm) quark mass denoted by $m_s\,(m_c)$ for $\Lambda\,(\Lambda_c)$ baryon. We adjust these parameters to fit the known masses  of  $\Lambda$ and $\Lambda_c$ compiled by the particle data group (PDG)~\cite{ParticleDataGroup:2020ssz}.
 Allowing an uncertainty on $\alpha_s$, we assimilate phenomenologically, in part, the effect of truncations on the system mass ($M$). 
\begin{table}[ht]
	\caption{The masses of $\Lambda$, $\Lambda_c$, and their isospin triplet baryons, i.e, $\Sigma^0$, $\Sigma^+$, $\Sigma^-$ and $\Sigma_c^0$, $\Sigma_c^+$, $\Sigma_c^{++}$ in units of MeV. Our results are compared with the experimental data~\cite{ParticleDataGroup:2020ssz}.}
	\centering 
		\begin{tabular}{|c|cc|}
			\hline \hline
			Baryons~~&~~  $M_{\rm BLFQ}$ ~~&~~  $M_{\rm exp}$ \\ 
			\hline 		
			
			$\Lambda$ ~~&~~  1116$^{+32}_{-48}$ ~~&~~  1115.683$\pm$0.006\\
			
			$\Sigma^0$ ~~&~~  1121$^{+37}_{-46}$  ~~&~~  1192.642$\pm$0.024\\
			
			$\Sigma^+$ ~~&~~   1120$^{+37}_{-46}$ ~~&~~  1189.37$\pm$0.07\\
			
			$\Sigma^-$ ~~&~~  1121$^{+37}_{-46}$ ~~&~~  1197.449$\pm$0.030\\
			\hline

			$\Lambda_c$ ~~&~~    2287$^{+7}_{-8}$ ~~&~~  2286.46$\pm$0.14\\ 
			
			$\Sigma_c^+$ ~~&~~    2290$^{+7}_{-7}$&  2452.9$\pm$0.4\\ 
			
			$\Sigma_c^{++}$ ~~&~~   2289$^{+7}_{-7}$ ~~&~~   2452.397$\pm$0.140\\ 
			
			$\Sigma_c^0$ ~~&~~   2291$^{+7}_{-8}$ ~~&~~  2452.375$\pm$0.140\\
			
			\hline \hline 
			
		\end{tabular}
	\label{table:masses}
\end{table}

Table~\ref{table:masses}  compares the computed masses for $\Lambda\,(\Sigma^0,\,\Sigma^+,\,\Sigma^-)$ and $\Lambda_c\,(\Sigma_c^+,\,\Sigma_c^{++},\,\Sigma_c^0$) in our BLFQ approach with the experimental data~\cite{ParticleDataGroup:2020ssz}. The errors appearing in our results are estimates based on our assigned $10\%$ uncertainty in the coupling constant $\alpha_s$. Note that we only fit the masses of $\Lambda$ and $\Lambda_c$, while the masses of their isospin states are our predictions. 
We find that the masses of these baryons are in the experimental range while isospin-dependent mass differences are small as compared to the experimental data. It should be noted that the current calculations subsume the gluon dynamics into effective interactions among the three valence quarks. We cannot directly access the dynamical role of the gluons due to the Fock space truncation. Future developments will focus on adding higher Fock sectors to include, for example, gluon and sea degrees of freedom, which will eventually allow us to incorporate the fundamental QCD interactions and provide a better prediction for isospin-dependent mass differences.

Using those model parameters given in Table~\ref{tab:parameter}, we then present the EMFFs and the PDFs of those baryons. We also predict the electromagnetic radii, and magnetic moments of those baryons.
\section{Electromagnetic form factors}\label{sec:EMFFs}
For spin-$1/2$ baryons, there are two independent EMFFs, namely the Dirac and the Pauli FFs, $F_1(Q^2)$ and $F_2(Q^2)$, respectively. In the light-front framework, they are identified with the helicity-conserving and helicity-flip matrix elements of the vector ($J^+\equiv\sum_q e_q \bar{\psi}_q\gamma^+\psi_q$) current:
\begin{align}	
	\left \langle  P+q,\uparrow \left | \frac{J^+(0)}{2P^+} \right |P, \uparrow\right \rangle =&F_1(Q^2)\,,\nonumber\\
	\left \langle  P+q,\uparrow \left | \frac{J^+(0)}{2P^+} \right |P, \downarrow\right \rangle =&-\frac{(q^1-iq^2)}{2M}\,F_2(Q^2)\,,\label{PFFs}
\end{align}

where $Q^2=-q^2$ is the square of the momentum transfer. Within the leading Fock-sector, the baryon state with momentum  $P$ and the light-front helicity $\Lambda$ can be expressed in terms of three-particle LFWFs:
\begin{align}
\ket{P,{\Lambda}} =&  \int \prod_{i=1}^{3} \left[\frac{{\rm d}x_i{\rm d}^2 \vec{p}_{i\perp}}{\sqrt{x_i}16\pi^3}\right] \nonumber\\
& \times 16\pi^3\delta \left(1-\sum_{i=1}^{3} x_i\right) \delta^2 \left(\sum_{i=1}^{3}\vec{p}_{i\perp}\right) \nonumber\\
& 
\times \Psi^{\Lambda}_{\{x_i,\vec{p}_{i\perp},\lambda_i\}} \ket{\{x_iP^+,\vec{p}_{i\perp}+x_i\vec{P}_{\perp},\lambda_i\}}\,,\label{wavefunction_expansion}
\end{align}

with $x_i=p_i^+/P^+$  and $\vec{p}_{i\perp}$ being the relative transverse momentum of the $i$-th quark. Substituting the baryonic states and the quark field operators ($\psi_q$ and $\bar{\psi}_q$) in Eq.~(\ref{PFFs}) provides the flavor Dirac and Pauli FFs in terms of the overlap of the LFWFs~\cite{Brodsky:2000xy}:
\begin{align}
F_1^q(Q^2)&= 
 \sum_{\lambda_i} \int \left[{\rm d}\mathcal{X} \,{\rm d}\mathcal{P}_\perp\right] \Psi^{\uparrow *}_{\{x^{\prime}_i,\vec{p}^{\prime}_{i\perp},\lambda_i\}}\Psi^{\uparrow}_{\{x_i,\vec{p}_{i\perp},\lambda_i\}} \,,    \\
F_2^q(Q^2)&= \frac{-(q^1-iq^2)}{2M}
 \sum_{\lambda_i} \int \left[{\rm d}\mathcal{X} \,{\rm d}\mathcal{P}_\perp\right]\nonumber\\&\quad\quad\quad\quad\times \Psi^{\uparrow *}_{\{x^{\prime}_i,\vec{p}^{\prime}_{i\perp},\lambda_i\}}\Psi^{\downarrow}_{\{x_i,\vec{p}_{i\perp},\lambda_i\}}\, , 
\end{align}
where $x^{\prime}_1=x_1$ and $\vec{p}^{\prime}_{1\perp}=\vec{p}_{1\perp}+(1-x_1)\vec{q}_{\perp}$ for the active quark, while $x^{\prime}_i={x_i}$ and $\vec{p}^{\prime}_{i\perp}=\vec{p}_{i\perp}-{x_i} \vec{q}_{\perp}$ for the spectators ($i=2,3$) and 
\begin{align}
\left[{\rm d}\mathcal{X} \,{\rm d}\mathcal{P}_\perp\right]=&\prod_{i=1}^3 \left[\frac{{\rm d}x_i{\rm d}^2 \vec{p}_{i\perp}}{16\pi^3}\right]\nonumber\\&\times 16 \pi^3 \delta \left(1-\sum_{i=1}^{3} x_i\right) \delta^2 \left(\sum_{i=1}^{3}\vec{p}_{i\perp}\right)\, .  
\end{align}
Here, we consider the frame where the momentum transfer is purely in the transverse direction, i.e., $q=(0,0,\vec{q}_{\perp})$, which implies $Q^2=-q^2={\vec{q}_{\perp}}^2$. 
 The FFs follow the normalizations $F_1^q(0)=n_q$, with $n_q$ being the number of valence quarks of flavor $q$ in the baryon, while the Pauli FFs at $Q^2=0$, provide the anomalous magnetic moments $F_2^q(0)=\kappa_q$.
\begin{figure*}[tph]
	\centering
	\includegraphics[width=0.7\textwidth]{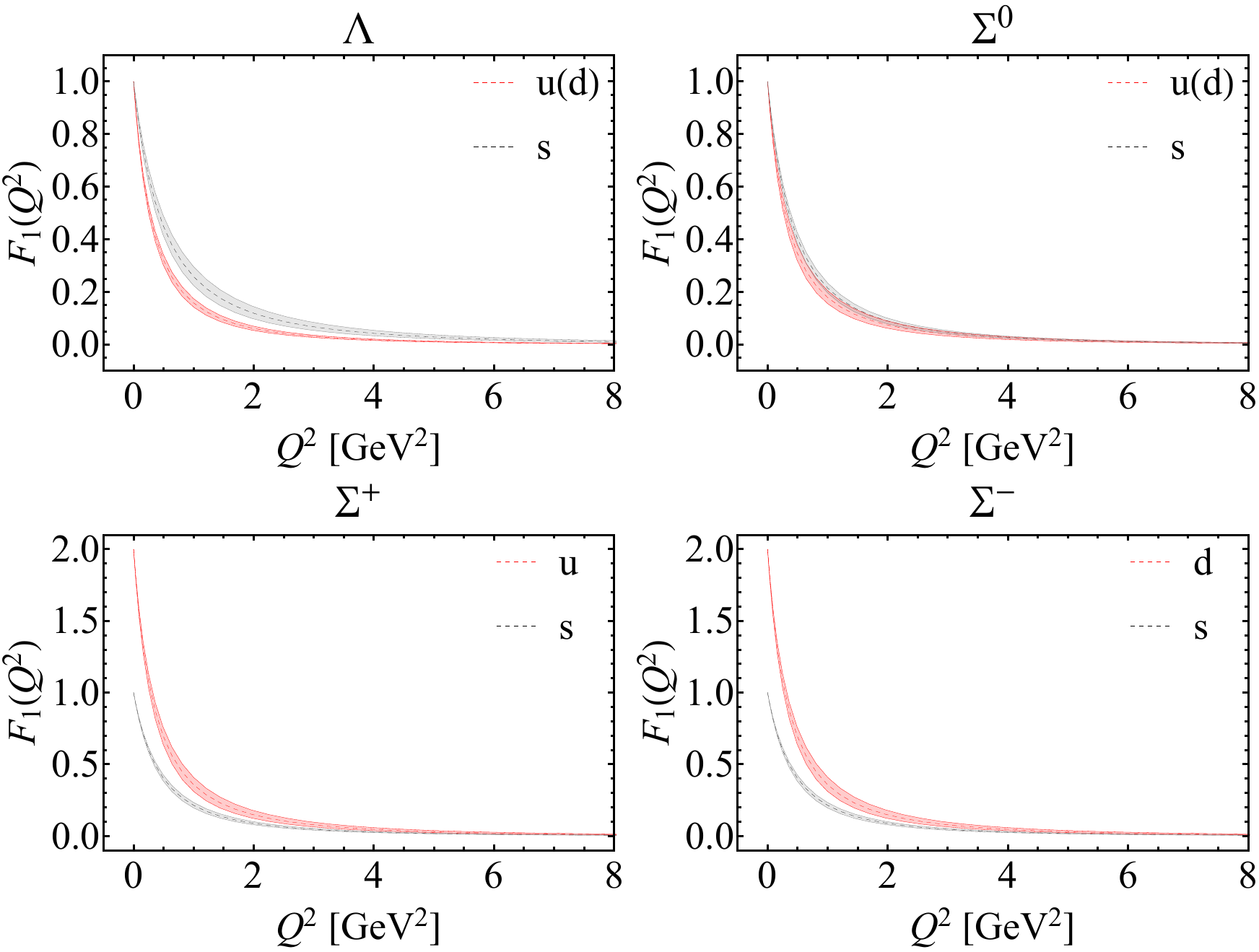}
	\caption{Flavor Dirac FFs of the $\Lambda$ baryon and its isospin states $(\Sigma^0,\Sigma^+,\Sigma^-)$. The red lines with red bands represent the light quark ($u$ and/or $d$) FFs, whereas the black lines with gray bands correspond to the strange quark ($s$) FFs. The bands reflect the $10\%$ uncertainty in the coupling constant $\alpha_s$.}
	\label{fig:lambda-ff1}
\end{figure*}
\begin{figure*}[tph]
	\centering
	\includegraphics[width=0.7\textwidth]{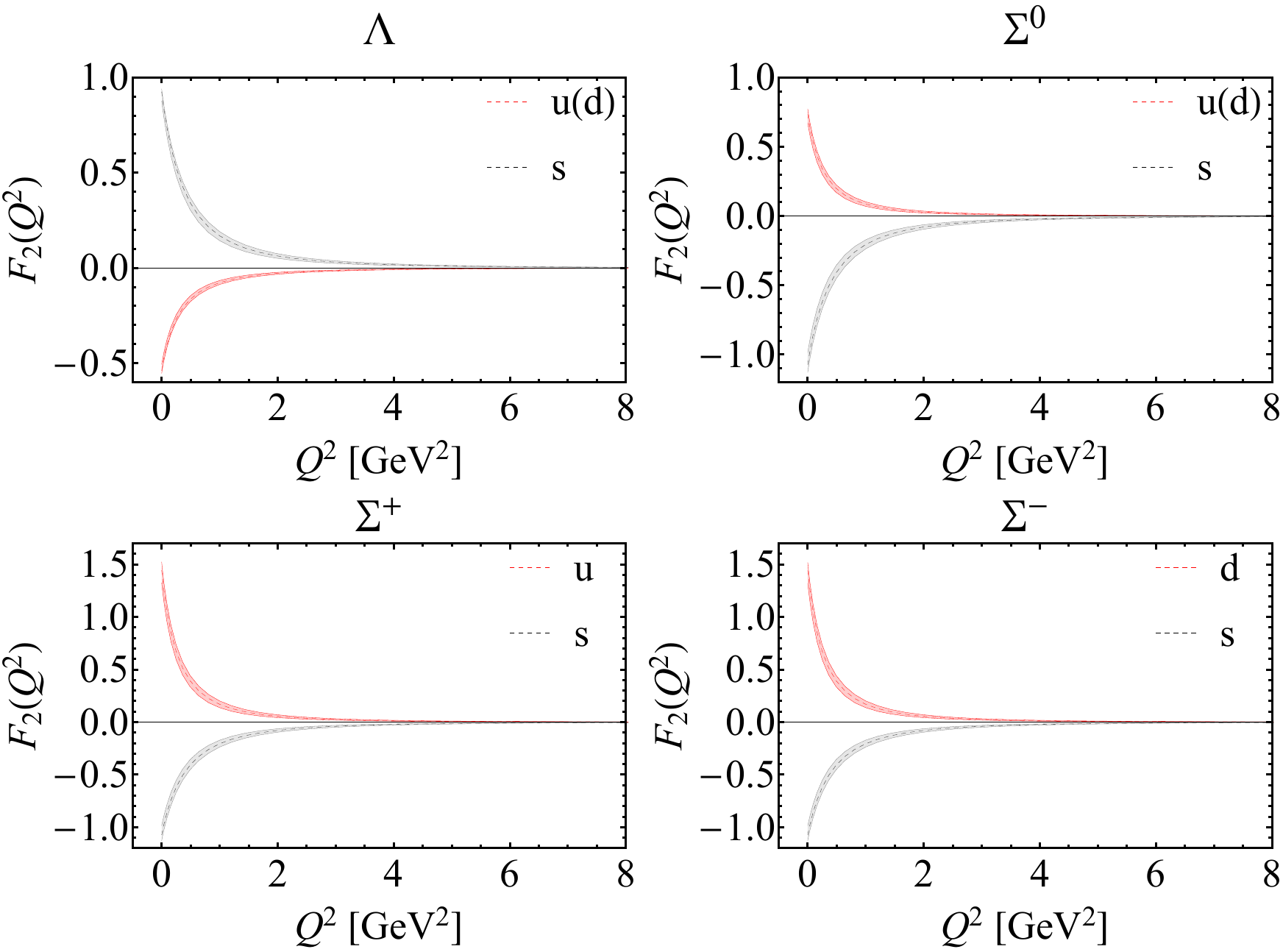}
	\caption{Flavor Pauli FFs of the $\Lambda$ baryon and its isospin states $(\Sigma^0,\Sigma^+,\Sigma^-)$. The red lines with red bands represent the light quark ($u$ and/or $d$) FFs, whereas the black lines with gray bands correspond to the strange quark ($s$) FFs. The bands reflect the $10\%$ uncertainty in the coupling constant $\alpha_s$.}
	\label{fig:lambda-ff2}
\end{figure*}
\begin{figure*}[tph]
    \includegraphics[width=0.35\textwidth]{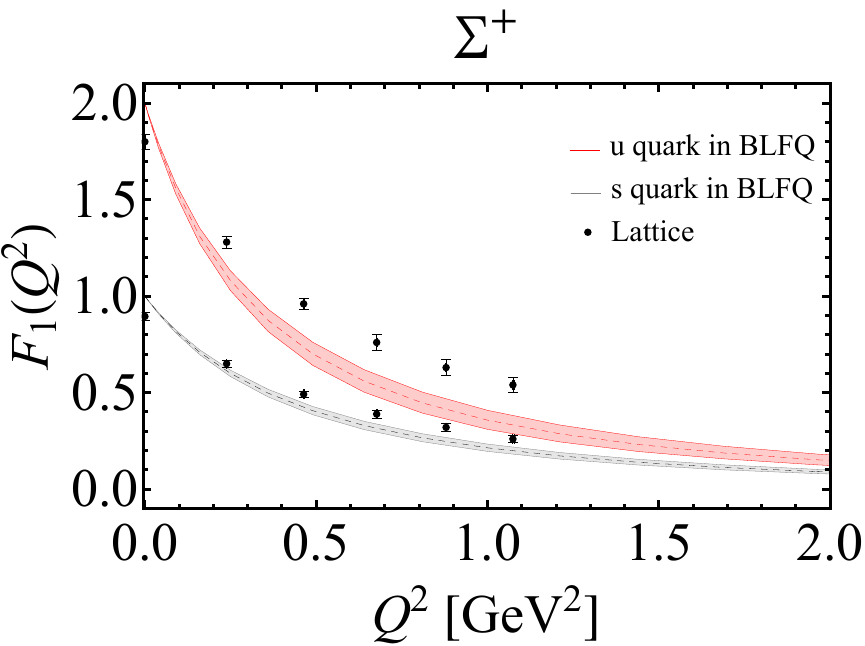}
    \includegraphics[width=0.35\textwidth]{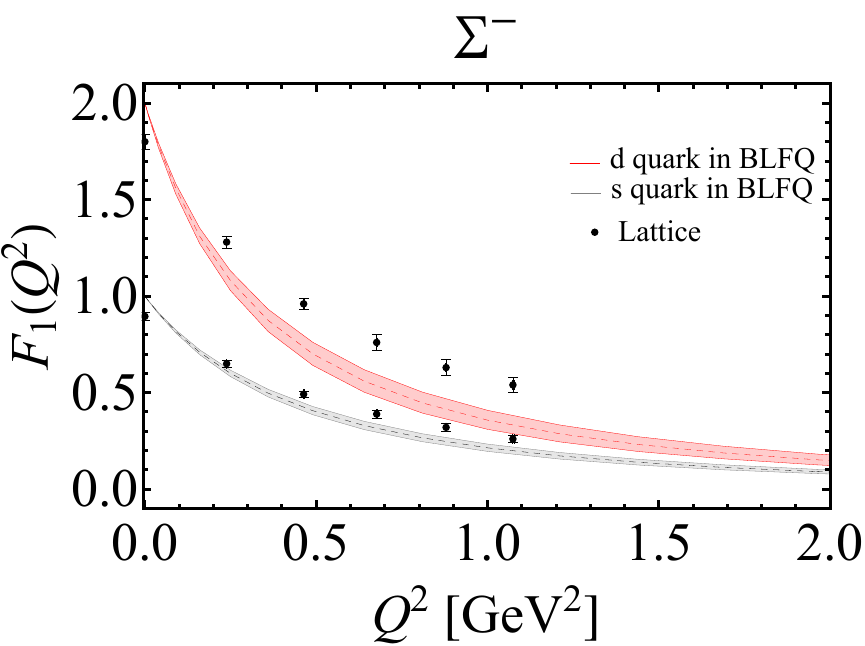}\\
    \includegraphics[width=0.35\textwidth]{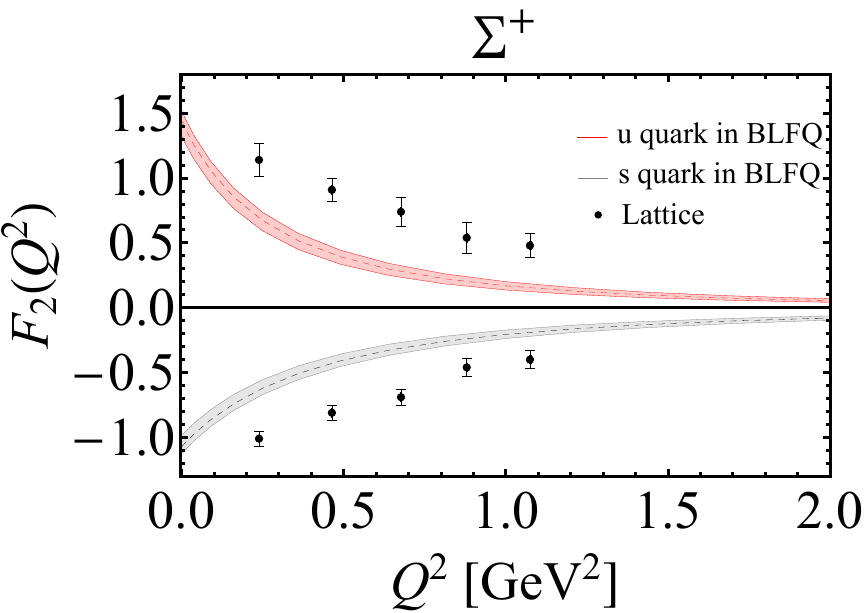}
	\includegraphics[width=0.35\textwidth]{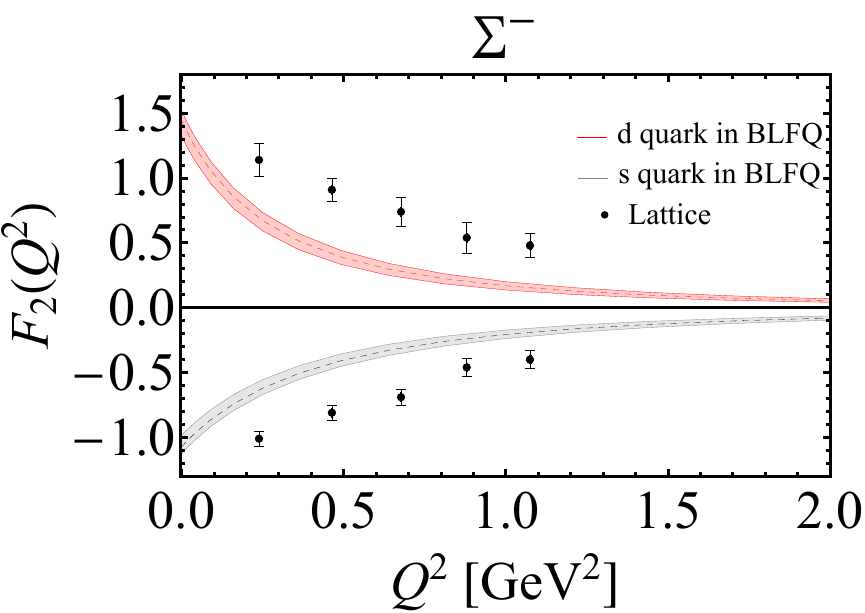}
	\caption{Comparison of the flavor FFs in $\Sigma^+$ and $\Sigma^-$ evaluated within BLFQ and the lattice QCD simulations~\cite{Lin:2008mr}. The red lines with red bands represent the light quark ($u$ and/or $d$) FFs, whereas the black lines with gray bands correspond to the strange quark ($s$) FFs.  The black pionts are lattice QCD results. The bands reflect the $10\%$ uncertainty in the coupling constant $\alpha_s$.}
	\label{fig:comparison_ff-Sigma}
\end{figure*}

We evaluate the Dirac and the Pauli FFs for the valence
quarks in the $\Lambda$, $\Lambda_c$, and their isospin states using the LFWFs defined in Eq.~(\ref{wavefunctions}). The flavor Dirac and Pauli 
FFs for $\Lambda$ and its isospin triplet states $(\Sigma^0,\Sigma^+,\Sigma^-)$ are shown in Figs.~\ref{fig:lambda-ff1} and \ref{fig:lambda-ff2}, respectively. The red bands represent the results for the light quark ($u$ and/or $d$), while the black bands correspond to the results for the $s$ quark. The error bands in our results are due to our adopted 10\% uncertainty in the coupling constant. The slope of the EMFF at $Q^2\to 0$ relates to the electromagnetic radius of the quark. We observe that at small $Q^2$ the slopes of the light quark FFs are larger than those of the $s$ quark FFs. This is due to the lighter mass of the up (down) quark compared to the $s$ quark, and thus the radius of the light quark is also larger than the $s$ quark. Although the flavor content is the same in $\Lambda$ and $\Sigma^0$,  their flavor FFs are not alike, while the flavor FFs of $\Sigma^+$ and $\Sigma^-$ are the same. The Pauli FF for the light quark in $\Lambda$ is negative but positive for the $s$ quark, whereas they exhibit opposite behavior in $\Sigma^0$.

In Fig.~\ref{fig:comparison_ff-Sigma}, we compare our results for the flavor FFs with the  lattice QCD simulations~\cite{Lin:2008mr} available only for the $\Sigma^+$ and $\Sigma^-$. We find that the qualitative behaviors of the flavor FFs obtained within our BLFQ approach and the lattice QCD simulations are approximately consistent with each other. It can be noticed from Figs.~\ref{fig:lambda-ff1}, \ref{fig:lambda-ff2}, \ref{fig:comparison_ff-Sigma} that our model preserves the isospin symmetry.

\begin{figure*}[tph]
	\includegraphics[width=0.7\textwidth]{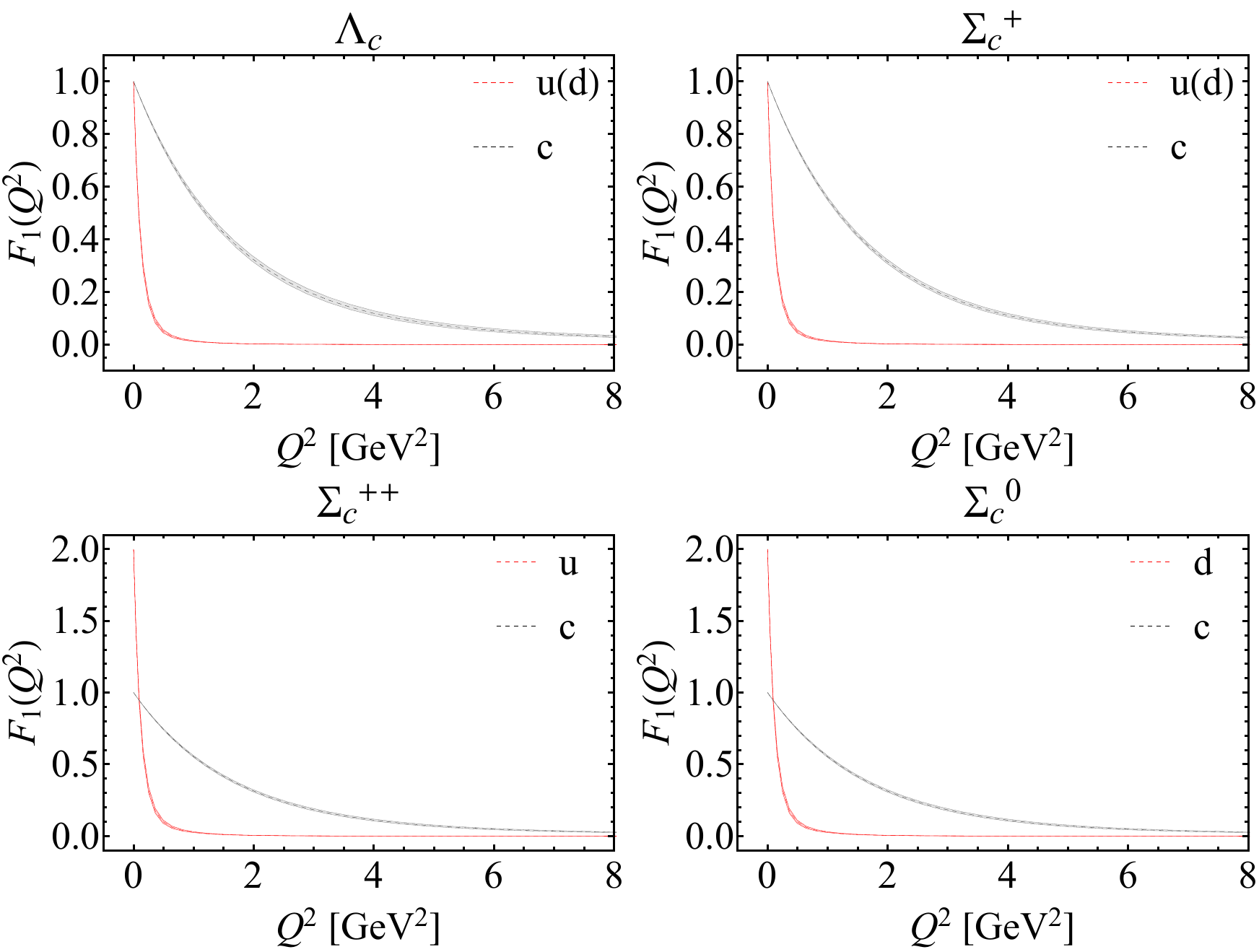}
	\caption{Flavor Dirac FFs of the $\Lambda_c$ baryon and its isospin states $(\Sigma^+_c,\Sigma^{++}_c,\Sigma^{0}_c)$. The red lines with red bands represent the light quark ($u$ and/or $d$) FFs, whereas the black lines with gray bands correspond to the charm quark ($c$) FFs. The bands reflect the $10\%$ uncertainty in the coupling constant $\alpha_s$.}
	\label{fig:lambdac-ff1}
\end{figure*}
\begin{figure*}[tph]
	\includegraphics[width=0.7\textwidth]{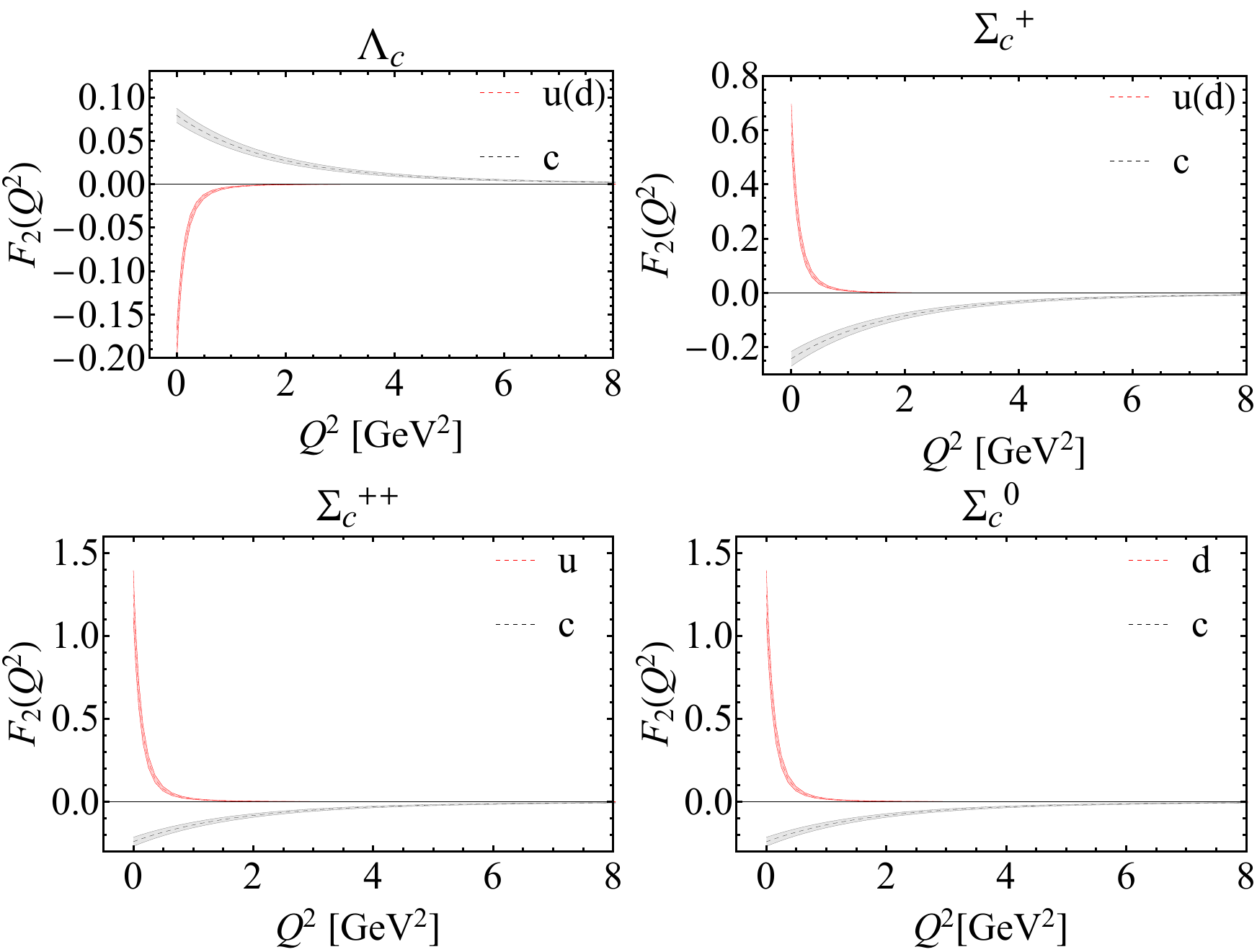}
	\caption{Flavor Pauli FFs of the $\Lambda_c$ baryon and its isospin states $(\Sigma^+_c,\Sigma^{++}_c,\Sigma^{0}_c)$. The red lines with red bands represent the light quark\textbf{} ($u$ and/or $d$) FFs, whereas the black lines with gray bands correspond to the charm quark ($c$) FFs. The bands reflect the $10\%$ uncertainty in the coupling constant $\alpha_s$.}
	\label{fig:lambdac-ff2}
\end{figure*}
\begin{figure*}[tph]
	\includegraphics[width=0.35\textwidth]{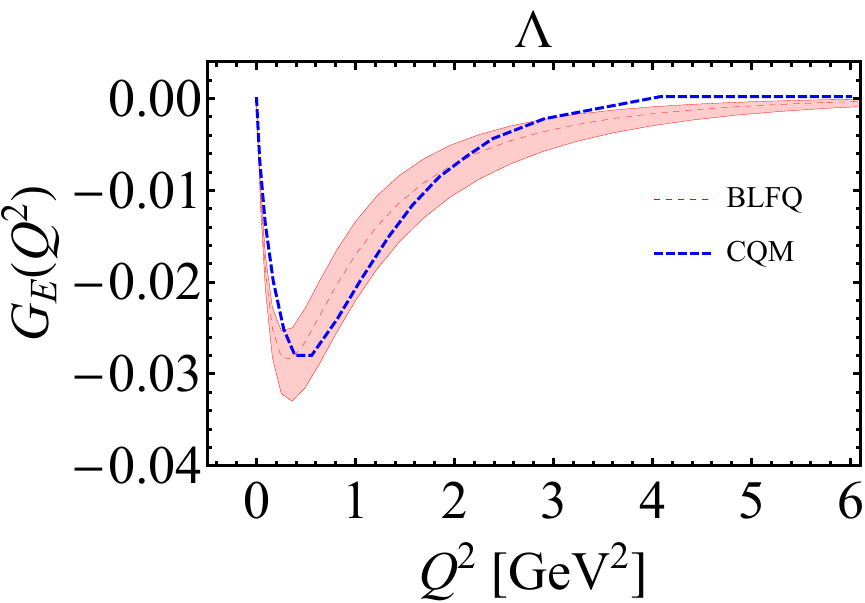}
	\includegraphics[width=0.35\textwidth]{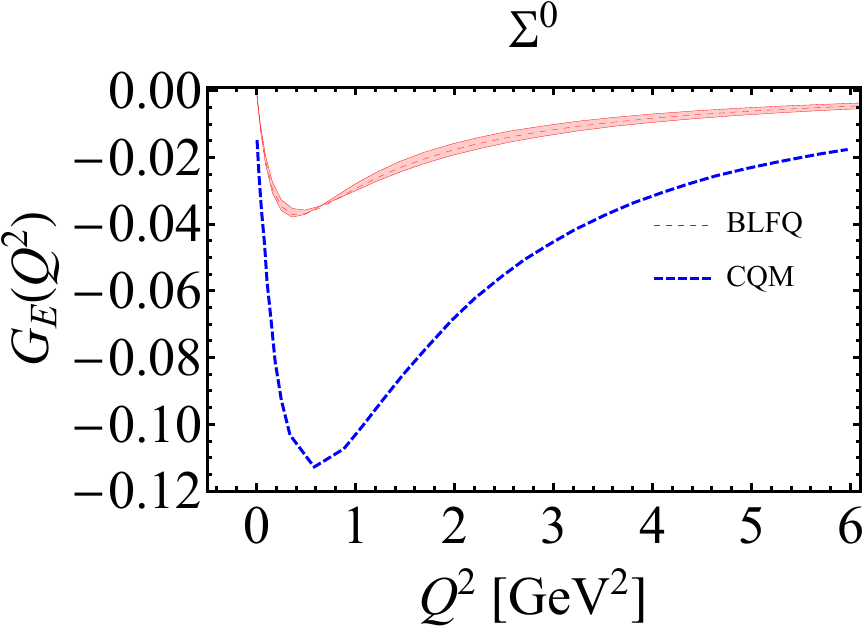}
    \includegraphics[width=0.35\textwidth]{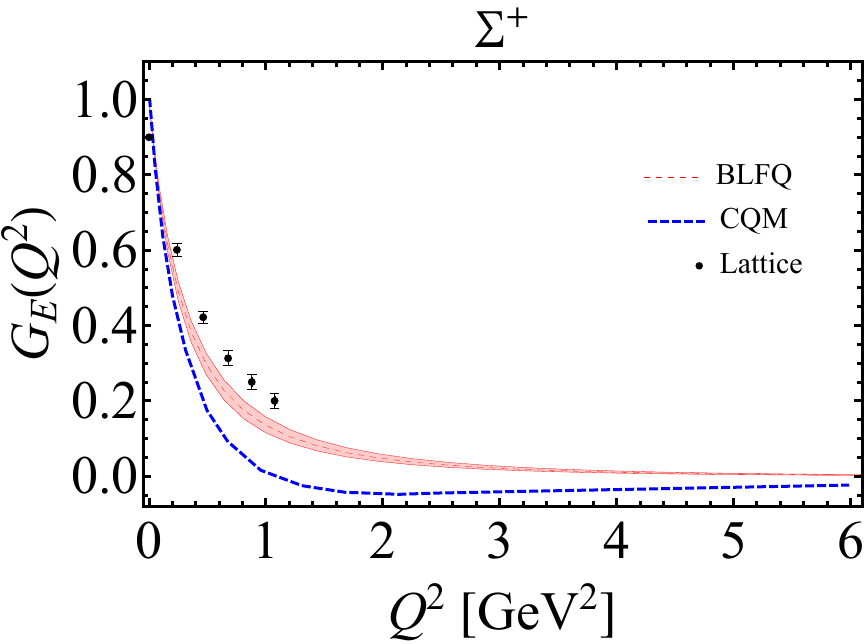}
    \includegraphics[width=0.35\textwidth]{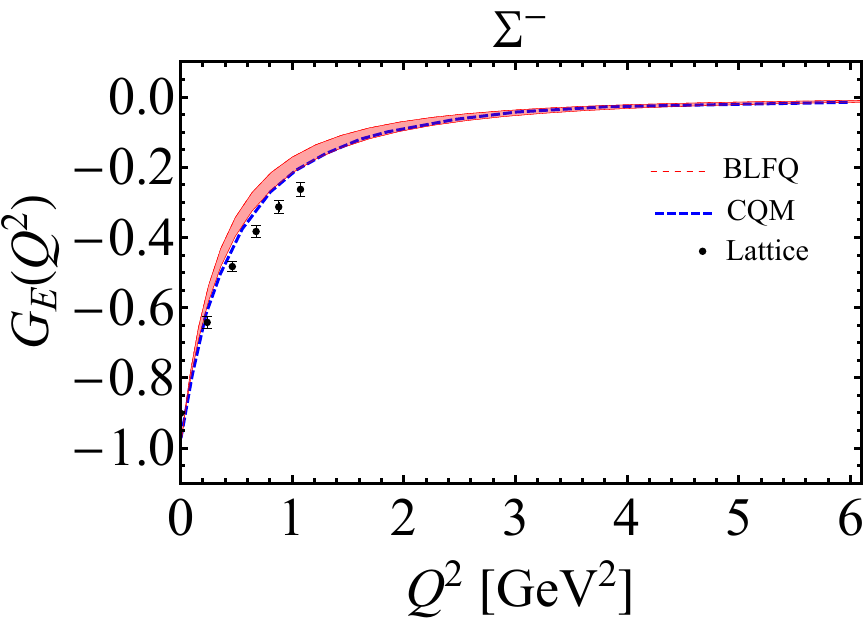}	
	\caption{Electric Sach's FFs $G_E(Q^2)$ as functions of $Q^2$ for $\Lambda(\Sigma^0,\Sigma^+,\Sigma^-)$. The red bands represent our results obtained within the BLFQ approach. The BLFQ results are compared with the lattice QCD simulations (black points)~\cite{Lin:2008mr}, and the CQM (dotted blue lines)~\cite{VanCauteren:2003hn}. The bands reflect the $10\%$ uncertainty in the coupling constant $\alpha_s$. }
	\label{fig:comparison_lambdaGE}
\end{figure*}

\begin{figure*}[tph]
	\includegraphics[width=0.35\textwidth]{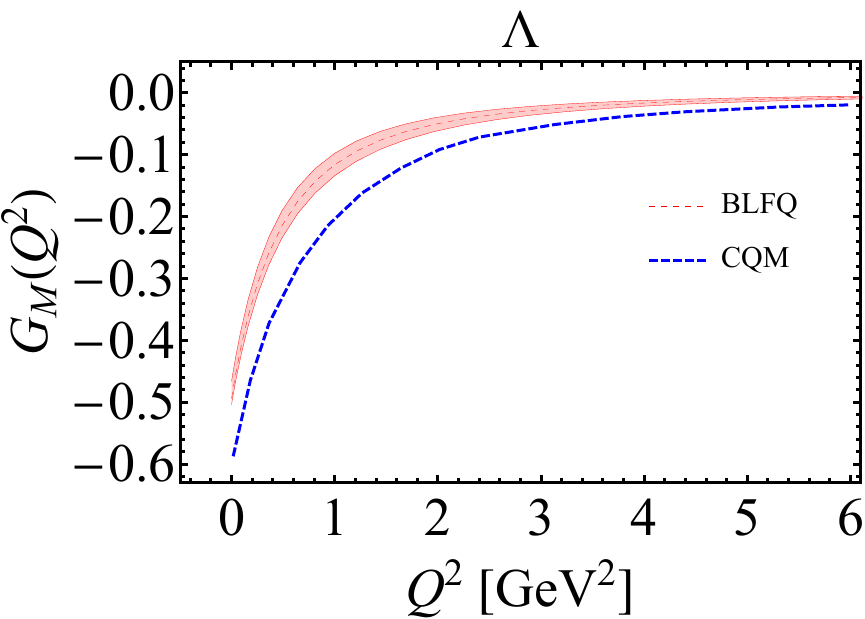}
	\includegraphics[width=0.35\textwidth]{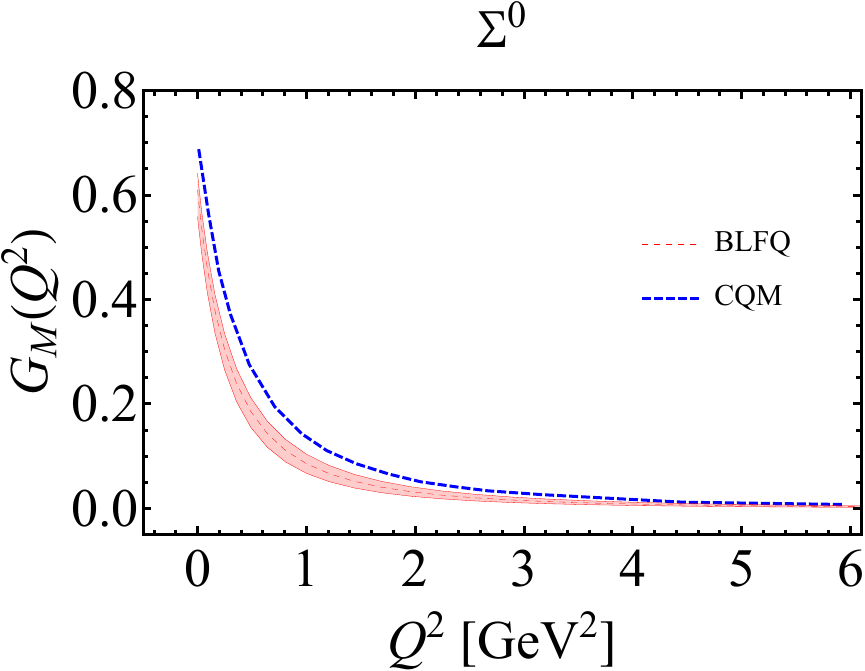}
	\includegraphics[width=0.35\textwidth]{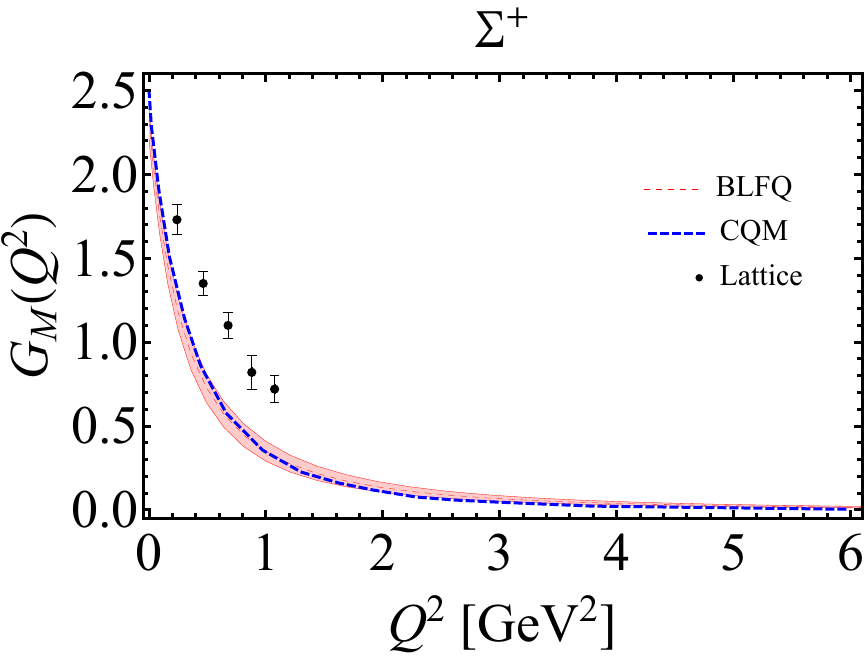}
	\includegraphics[width=0.35\textwidth]{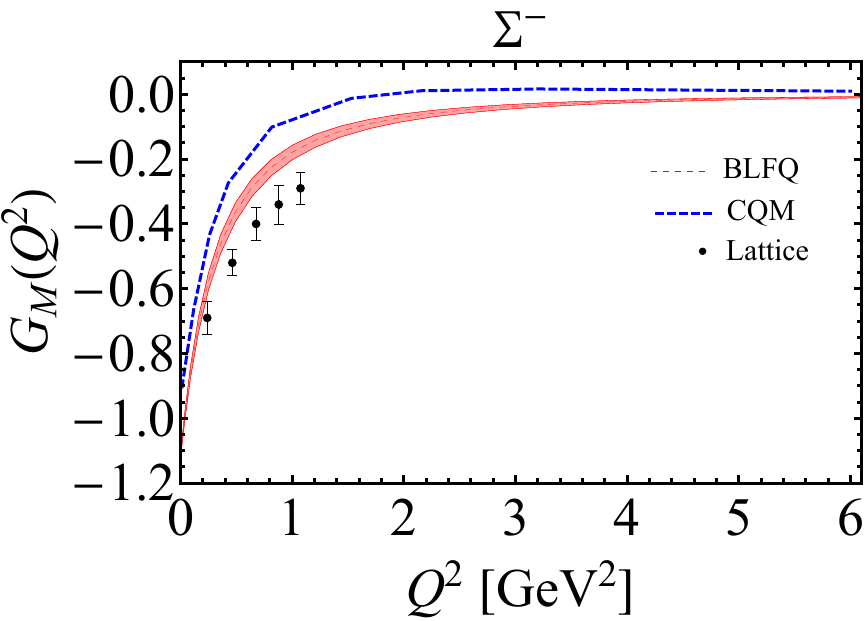}	
	\caption{Magnetic Sach's FFs $G_M(Q^2)$ as functions of $Q^2$ for $\Lambda(\Sigma^0,\Sigma^+,\Sigma^-)$. The gray bands represent our results obtained within the BLFQ approach. The BLFQ results are compared with the lattice QCD simulations (black points)~\cite{Lin:2008mr}, and the CQM (dotted blue lines)~\cite{VanCauteren:2003hn}. The bands reflect the $10\%$ uncertainty in the coupling constant $\alpha_s$. }
	\label{fig:comparison_lambdaGM}
\end{figure*}

Figures~\ref{fig:lambdac-ff1} and \ref{fig:lambdac-ff2} present the flavor Dirac and Pauli FFs, respectively, for the $\Lambda_c$ baryon and its isospin states $(\Sigma^+_c,\Sigma^{++}_c,\Sigma^{0}_c)$. Here again, we observe that the light quark FFs fall much faster than that of the $c$ quark indicating that, as may be expected, the $c$ quark is more localized near the center of the baryons than the light quark. The flavor Dirac FFs of the $\Lambda_c$ and $\Sigma^+_c$ are found to be alike but their flavor Pauli FFs change sign. Meanwhile, both the flavor FFs of $\Sigma^{++}_c$ are identical to that of $\Sigma^{0}_c$.

Under charge and isospin symmetry, the baryon FFs
can be obtained from the flavor FFs
 \begin{align}
 F_{1(2)}^{\rm B}(Q^2)=\sum_q e_q F_{1(2)}^q\,,
 \end{align} 
 where the charges of the quarks $e_u=\frac{2}{3}$, $e_d=-\frac{1}{3}$,  $e_s=-\frac{1}{3}$, and $e_c=\frac{2}{3}$.
Meanwhile, the Sachs FFs are expressed in terms of Dirac and Pauli FFs as
\begin{align}\label{GEGM}
 G_E^{\rm B}(Q^2)&= F_1^{\rm B}(Q^2)-\frac{Q^2}{4M^2}F_2^{\rm B}(Q^2),\nonumber\\
 G_M^{\rm B}(Q^2)&= F_1^{\rm B}(Q^2)+F_2^{\rm B}(Q^2)\,.
\end{align}

\begin{figure*}[tph]
	\includegraphics[width=0.7\textwidth]{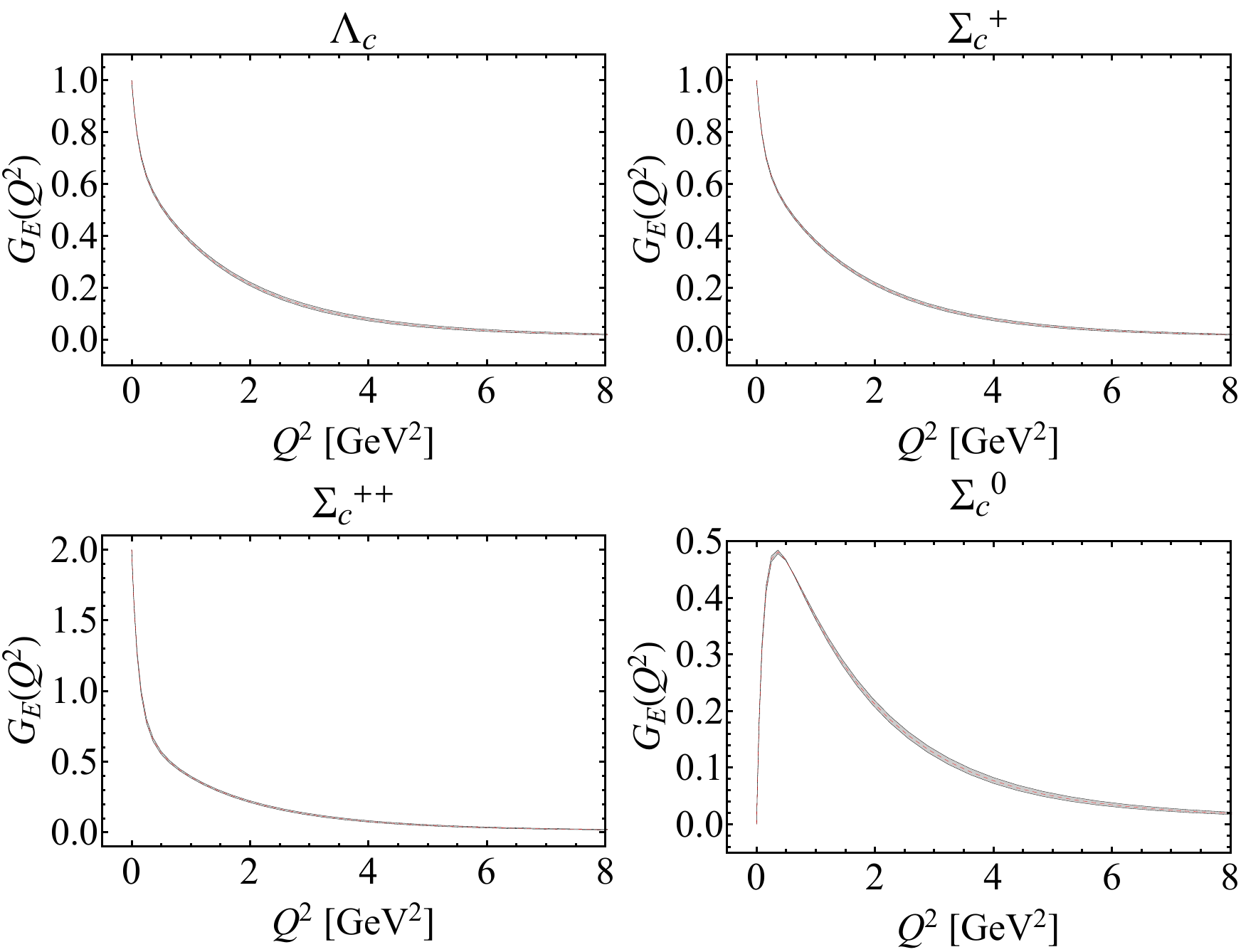}
	\caption{Electric Sach's FFs $G_E(Q^2)$ as functions of $Q^2$ for $\Lambda_c(\Sigma^0_c,\Sigma^+_c,\Sigma^-_c)$. The bands represent our results obtained within the BLFQ approach.  The bands reflect the $10\%$ uncertainty in the coupling constant $\alpha_s$.}
	\label{fig:lambdac-GE}
\end{figure*}
\begin{figure*}[tph]
	\includegraphics[width=0.7\textwidth]{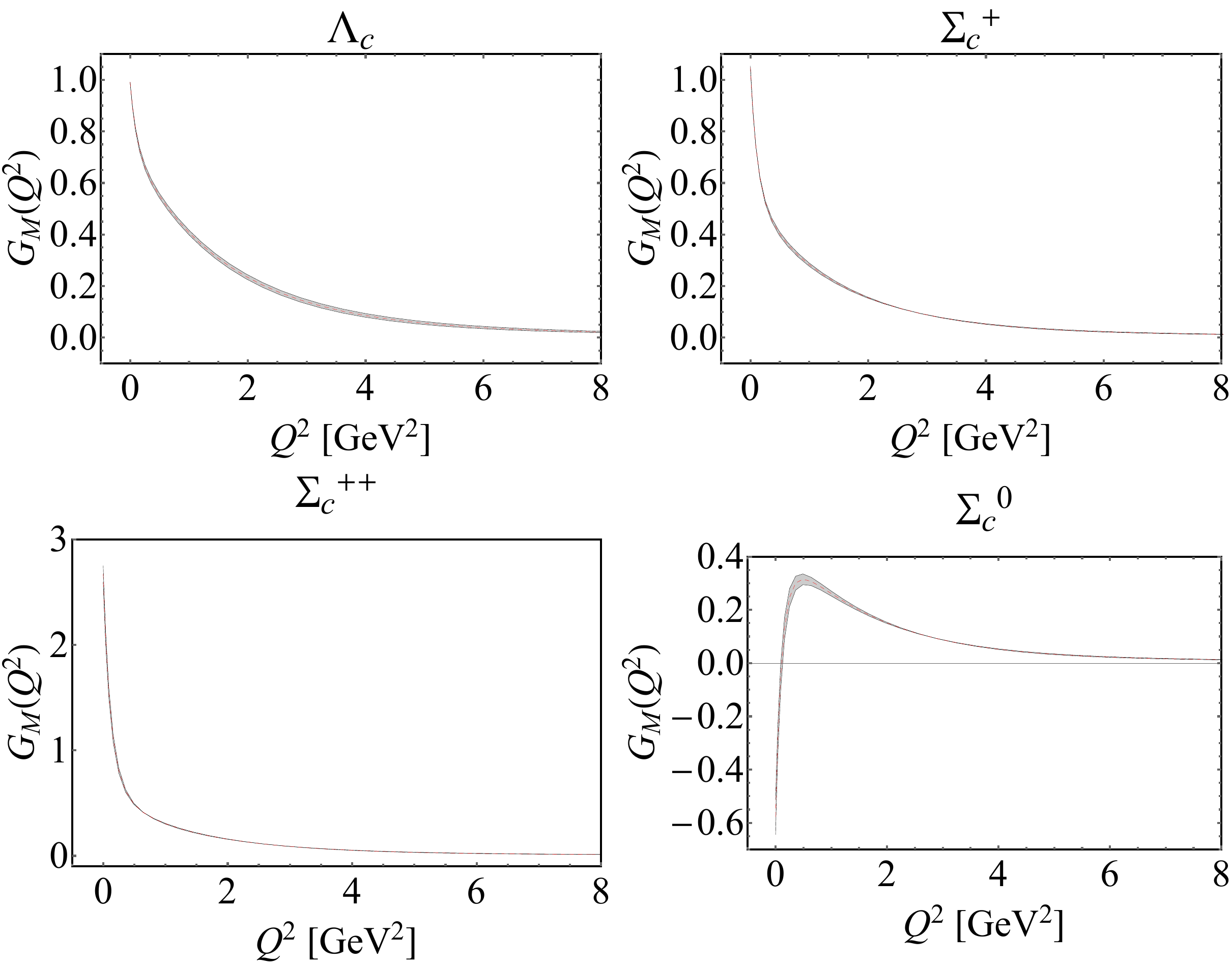}
	\caption{Magnetic Sach's FFs $G_M(Q^2)$ as functions of $Q^2$ for $\Lambda_c(\Sigma^0_c,\Sigma^+_c,\Sigma^-_c)$. The bands represent our results obtained within the BLFQ approach.  The bands reflect the $10\%$ uncertainty in the coupling constant $\alpha_s$.}
	\label{fig:lambdac-GM}
\end{figure*}

We show the electric and magnetic Sachs FFs of the $\Lambda$ and its isospin triplet states in Figs.~\ref{fig:comparison_lambdaGE} and \ref{fig:comparison_lambdaGM}, respectively. We compare our BLFQ results for all states with the results computed in the Constituent Quark Model (CQM)~\cite{VanCauteren:2003hn}. The FFs for $\Sigma^+$ and $\Sigma^-$ are also compared with available lattice QCD simulations~\cite{Lin:2008mr}. Qualitatively, our results are consistent with those theoretical calculations~\cite{Lin:2008mr,VanCauteren:2003hn}. The Sachs FFs for the $\Lambda_c$ and its isospin triplet states are presented in Figs.~\ref{fig:lambdac-GE} and \ref{fig:lambdac-GM}.

The magnetic moments of the baryons are related to the baryons' magnetic Sachs FFs at $Q^2=0$. 
In our approach, we obtain the magnetic moment of the $\Lambda$, $\mu_{\rm{\Lambda}}=-0.494^{+0.028}_{-0.010}$ and for the isospin states, $\mu_{\rm{\Sigma^0}}= 0.610^{+0.032}_{-0.051}$, $\mu_{\rm{\Sigma^+}}= 0.2323^{+0.067}_{-0.112}$, and $\mu_{\rm{\Sigma^-}}= -1.124^{+0.011}_{-0.007}$ close to the available measurements~\cite{GoughEschrich:1998tx} for $\Lambda$, $\Sigma^+$ and $\Sigma^-$. Note that $\mu_{\Sigma^0}$ has not been measured. However, it is given by $\mu_{\Sigma^0}=(\mu_{\Sigma^+}+\mu_{\Sigma^-})/2$ according to the isospin symmetry. The magnetic moments of the baryons are compared with experimental data in Table~\ref{table-lambda-magneticmoment}. Based on our BLFQ approach, the magnetic moments for
$\Lambda_c$ and its isospin triplet states are given in Table~\ref{table-lambda_c-magneticmoment}, where we compare our results with other theoretical calculations in Refs.~\cite{Julia-Diaz:2004yqv,Faessler:2006ft,Sharma:2010vv,Barik:1983ics,Bernotas:2012nz,Zhu:1997as,Kumar:2005ei,Patel:2007gx,Yang:2018uoj,Wang:2018gpl,Wang:2018gpl}.
We find that our predictions for the magnetic
moments of $\Lambda_c$ and $\Sigma_c^+$ are larger than those in Refs.~\cite{Julia-Diaz:2004yqv,Faessler:2006ft,Sharma:2010vv,Barik:1983ics,Bernotas:2012nz,Zhu:1997as,Kumar:2005ei,Patel:2007gx,Yang:2018uoj,Wang:2018gpl,Wang:2018gpl}. This is attributed to the anomalous magnetic moment, i.e. $F_2(0)$, contribution from the charm quark. In our calculations, we find that $F_2(0)$ for the charm quark is negative and significantly smaller than the light quarks, while in other theoretical approaches, the magnitude of $F_2^c(0)$ is much larger than our result and comparable with their $F_2(0)$ of light quarks. Consequently,  the anomalous magnetic moments of $\Lambda_c$ and $\Sigma_c^+$, $F_2^B(0)=\sum_q e_q F_2^q(0)$, in our approach are larger than other theoretical predictions. On the other hand, our result for the magnetic
moment of $\Sigma_c^{++}$ roughly agrees with those in Refs.~\cite{Julia-Diaz:2004yqv,Sharma:2010vv,Barik:1983ics,Zhu:1997as,Kumar:2005ei,Patel:2007gx,Yang:2018uoj} and is higher than the predictions reported in Refs.~\cite{Faessler:2006ft,Bernotas:2012nz,Wang:2018gpl}. Meanwhile, our predicted value for the magnetic moments of $\Sigma_c^0$ is lower than the values outlined in Refs.~\cite{Julia-Diaz:2004yqv,Faessler:2006ft,Sharma:2010vv,Barik:1983ics,Bernotas:2012nz,Zhu:1997as,Kumar:2005ei,Patel:2007gx,Yang:2018uoj,Wang:2018gpl,Wang:2018gpl}. Note that nothing is known about the magnetic moment of $\Lambda_c$ and its isospin triplet states experimentally at at the present time.. Our calculation provides a prediction of the expected data for the magnetic moments of heavy baryon from the
future experiments.

\begin{table}[ht]
	\caption{The magnetic moments of $\Lambda(\Sigma^0,\,\Sigma^+,\,\Sigma^-)$. Our results are compared with the experimental data \cite{ParticleDataGroup:2020ssz} (in units of the nuclear magneton $\mu_{\rm N}$).}
	\centering 
		\begin{tabular}{|c|cc|}
			\hline \hline
		Baryons	~&~ $\mu_{\rm BLFQ} $ ~~~&~~~ $\mu_{\rm exp}$\cite{ParticleDataGroup:2020ssz} \\ 
			\hline 		
			
			$\Lambda$&-0.494$^{+0.028}_{-0.010}$&  -0.613$\pm$0.004\\
			
			$\Sigma^0$& 0.610$^{+0.032}_{-0.051}$&  ---\\
			
			$\Sigma^+$& 2.323$^{+0.067}_{-0.112}$&  2.458$\pm$0.010\\
			
			$\Sigma^-$& -1.124$^{+0.011}_{-0.007}$&  -1.160$\pm$0.025\\
			
			\hline \hline 
			
		\end{tabular}
	\label{table-lambda-magneticmoment}
\end{table}
\begin{table*}[ht]
	\caption{Our predictions for the magnetic moments of $\Lambda_c(\Sigma^+_c,\,\Sigma^{++}_c,\,\Sigma^0_c)$ (in units of the nuclear magneton $\mu_{\rm N}$). Our results are compared with other theoretical calculations in Refs.~\cite{Julia-Diaz:2004yqv,Faessler:2006ft,Sharma:2010vv,Barik:1983ics,Bernotas:2012nz,Zhu:1997as,Kumar:2005ei,Patel:2007gx,Yang:2018uoj,Wang:2018gpl,Wang:2018gpl}. In Ref.~\cite{Wang:2018gpl}, S-I and S-II represent the results obtained using two different sets of parameters.}
	\centering 
		\begin{tabular}{|c|c|ccccccccccc|}
			\hline \hline
			Baryons ~~&~~ $\mu_{\rm BLFQ} $                         ~~~&~~~\cite{Julia-Diaz:2004yqv}    ~&~\cite{Faessler:2006ft}     ~&~\cite{Sharma:2010vv} ~&~\cite{Barik:1983ics}  ~&~\cite{Bernotas:2012nz} ~&~\cite{Zhu:1997as} ~&~\cite{Kumar:2005ei} ~&~\cite{Patel:2007gx}  ~&~\cite{Yang:2018uoj}~&~S-I~\cite{Wang:2018gpl}~&~S-II~\cite{Wang:2018gpl} \\ 
			\hline 	
			
			$\Lambda_c$& 0.99$^{+0.00}_{-0.00}$   & 0.41 &0.42 &0.392 &0.341 &0.411 &--- &0.37 &0.385 &--- &0.24&0.24\\ 
			
			$\Sigma_c^+$& 1.05$^{+0.01}_{-0.01}$     &0.65 &0.36 &0.30 &0.525 &0.318 &--- &0.63 &0.501 &0.46(3)&0.26&0.30\\ 
			
			$\Sigma_c^{++}$& 2.67$^{+0.49}_{-0.08}$     &3.07 &1.76 &2.20 &2.44 &1.679 &2.1(3) &2.18 &2.279 &2.15(10)&1.50&1.50\\ 
			
			$\Sigma_c^0$& -0.58$^{+0.06}_{-0.07}$  &-1.78 &-1.04 &-1.60 &-1.391 &-1.043 &-1.6(2) &-1.17 &-1.015 &-1.24(5)&-0.97&-0.91\\
			
			\hline \hline 
		\end{tabular} 
	\label{table-lambda_c-magneticmoment} 
\end{table*}

\newpage
From the Sachs FFs, we can also compute the electromagnetic radii of the baryons (When the charge of the baryon is zero, we need to replace the $G_{\rm E}^{\rm B}(0)$ with 1), which are defined by
\begin{align}
\langle r^2_{\rm E}\rangle^{\rm B}=&-\frac{6}{G_{\rm E}^{\rm B}(0)} \frac{{\rm d}G_{\rm E}^{\rm B}(Q^2)}{{\rm d}Q^2}\bigg|_{Q^2=0}, \\
\langle r^2_{\rm M} \rangle^{\rm B}=&-\frac{6}{G_{\rm M}^{\rm B}(0)}\frac{{\rm d}G_{\rm M}^{\rm B}(Q^2)}{{\rm d}Q^2}\bigg|_{Q^2=0}.
\end{align}
The radii of $\Lambda(\Sigma^0,\,\Sigma^+,\,\Sigma^-)$ are presented in Tables~\ref{table-lambda-charge-radius} and \ref{table-lambda-magnetic-radius}. We compare the BLFQ results with the available theoretical calculations~\cite{Kubis:2000aa,Puglia:2000jy,Julia-Diaz:2004yqv} and the only available measured data for the charge radius of $\Sigma^-$~\cite{GoughEschrich:1998tx}. We find a reasonable agreement with the experiment within our 10\% uncertainties stemming from our uncertainty in $\alpha_s$. We also find reasonable consistency between our predictions and the results evaluated  in the framework of heavy baryon chiral perturbation theory~\cite{Kubis:2000aa}.

 Our predictions for the electromagnetic radii of $\Lambda_c(\Sigma^+_c,\,\Sigma^{++}_c,\,\Sigma^0_c)$ are given in Tables~\ref{table-lambda_c-charge radii} and \ref{table-lambda_c-magnetic radii}. The charge radii are compared with the results evaluated  in  relativistic quark models~\cite{Julia-Diaz:2004yqv}. Here we observe substantial differences  between our BLFQ predictions and the relativistic quark models~\cite{Julia-Diaz:2004yqv}. In this connection, we note that there is a rather large spread in the results of the relativistic quark models.

\begin{table*}[ht]
	\caption{Our predictions for the charge radius $\langle r_{\rm E}^2\rangle$ for $\Lambda(\Sigma^0,\,\Sigma^+,\,\Sigma^-)$ in the unit of fm$^2$. Our results are compared with the heavy baryon chiral perturbation theory (HB$\chi$PT)~\cite{Kubis:2000aa,Puglia:2000jy}, the relativistic quark models (RQM)~\cite{Julia-Diaz:2004yqv} and the experimental data~\cite{GoughEschrich:1998tx} available only for $\Sigma^-$. In Ref.~\cite{Kubis:2000aa}, the results were computed with different regularization procedures: heavy-baryon (HB) approach and infrared regularization (IR) scheme.}
	\centering 
		\begin{tabular}{|c|c|c|c|c|c|c|c|c|c|c|}
			\hline \hline
			& $\langle r_{\rm E}^2\rangle_{\rm BLFQ}$  & \multicolumn{2}{c|}{HB~\cite{Kubis:2000aa}}  &  \multicolumn{2}{c|}{IR~\cite{Kubis:2000aa}} &  \multicolumn{2}{c|}{HB$\chi$PT~\cite{Puglia:2000jy}}  &\multicolumn{2}{c|}{RQM~\cite{Julia-Diaz:2004yqv}} & experimental data~\cite{GoughEschrich:1998tx}\\ \cline{3-10}  
			
			&       & $O(q^3)$&$O(q^4)$  & $O(q^3)$&$O(q^4)$  &$O(1/\Lambda_{\chi}^2) $ &$O(1/\Lambda_{\chi}^2M_N) $ &I&II& \\ 
			\hline 		
			$\Lambda$   &  0.07$\pm0.01$ &   0.14 &0.00  &   0.05&$0.11\pm0.02$  &  -0.150&  -0.050 &-0.01&0.02&      ---          \\
			
			$\Sigma^0$  &  0.07$^{+0.00}_{-0.01}$  &   -0.14 &-0.08 &   -0.05&$-0.03\pm0.01$ & ---  & ---  &0.02&0.02&       ---           \\
			
			$\Sigma^+$  &  0.79$\pm0.05$  &   0.59 &0.72  &   0.63&$0.60\pm0.02$   &1.522 &1.366  &0.47&0.66 &       ---           \\
			
			$\Sigma^-$  &  0.65$\pm0.02$ &   0.87 &0.88  &   0.72&$0.67\pm0.03$  &0.977 &0.798   &0.41&0.64& 0.60$\pm$0.08$\pm$0.08     \\
			
			\hline \hline 
			
		\end{tabular}
	\label{table-lambda-charge-radius}
\end{table*}


\begin{table}[ht]
	\caption{Our predictions for the magnetic radius $\langle r_{\rm M}^2\rangle$ for $\Lambda(\Sigma^0,\,\Sigma^+,\,\Sigma^-)$ in the unit of fm$^2$. Our results are compared with other theoretical calculations  in the framework of heavy baryon chiral
perturbation theory in Refs.~\cite{Kubis:2000aa}.}
	\centering 
		\begin{tabular}{|c|ccc|}
			\hline \hline
		Baryons	& $\langle r_{\rm M}^2\rangle_{\rm BLFQ}$                                & $O(q^4)$ HB\cite{Kubis:2000aa}        &  $O(q^4)$ IR\cite{Kubis:2000aa}\\ 
			\hline 		
			
			$\Lambda$&   0.52$\pm0.01$   &  $0.30\pm0.11$    &  $0.48\pm0.09$\\
			
			$\Sigma^0$&   0.82$^{+0.00}_{-0.01}$  & $0.20\pm0.10$     &  $0.45\pm0.08$\\
			
			$\Sigma^+$&   0.79$\pm0.00$   & $0.74\pm0.06$     &  $0.80\pm0.05$\\
			
			$\Sigma^-$&   0.70$\pm0.02$  & $1.33\pm0.16$     &  $1.20\pm0.13$\\
			
			\hline \hline 
			
		\end{tabular}
	\label{table-lambda-magnetic-radius}
\end{table}
\begin{table}[ht]
	\caption{Our predictions for the charge radius $\langle r_{\rm E}^2\rangle$ for $\Lambda_c(\Sigma^+_c,\,\Sigma^{++}_c,\,\Sigma^0_c)$ in the unit of fm$^2$. Our results are compared with the results in relativistic quark models~\cite{Julia-Diaz:2004yqv}. The “instant”, “point”, and “front” are three forms of kinematics, first outlined by Dirac~\cite{Dirac:1949cp}.}
	\centering 
		\begin{tabular}{|c|c|c|c|c|}
			\hline \hline
			Baryons &  $\langle r_{\rm E}^2\rangle_{\rm BLFQ}$ &\multicolumn{3}{c|}{Ref.~\cite{Julia-Diaz:2004yqv}} \\ \cline{3-5} 
			
			&                            &Instant&Point&Front \\  
			\hline 	
			
			$\Lambda_c$&  0.73$^{+0.02}_{-0.02}$     &0.5 &0.2 &0.4\\ 			
			$\Sigma_c^+$&  0.74$^{+0.02}_{-0.02}$    &0.5 &0.2 &0.4\\ 			
			$\Sigma_c^{++}$&  1.33$^{+0.03}_{-0.03}$ &1.7 &0.4 &1.4\\ 		
			$\Sigma_c^0$& $-1.19$ $^{0.039}_{-0.03}$    & $-0.7$ & $-0.0$ &$-0.6$\\
			
			\hline \hline 
		\end{tabular} 
	\label{table-lambda_c-charge radii} 
\end{table}
\begin{table}[ht]
	\caption{Our predictions for the magnetic radii $\langle r_{\rm M}^2\rangle$ for $\Lambda_c(\Sigma^+_c,\,\Sigma^{++}_c,\,\Sigma^0_c)$ in the unit of fm$^2$.}
	\centering 
		\begin{tabular}{|c|c|}
			\hline \hline
			Baryons &     $\langle r_{\rm M}^2\rangle_{\rm BLFQ}$ \\ 
			\hline 	
			
			$\Lambda_c$& 0.64$^{+0.03}_{-0.03}$\\ 
			
			$\Sigma_c^+$&  0.78$^{+0.01}_{-0.01}$\\ 
			
			$\Sigma_c^{++}$&  1.54$^{+0.01}_{-0.01}$\\ 
			
			$\Sigma_c^0$&   3.37$^{+0.33}_{-0.27}$\\
			
			\hline \hline 
		\end{tabular} 
	\label{table-lambda_c-magnetic radii} 
\end{table}

\section{Parton distribution functions of the baryons}\label{sec:pdfs}

The PDF, the probability density that a parton carries a certain fraction of the total light-front longitudinal momentum of a hadron, provides us with information about the nonperturbative structure of hadrons. The quark PDF of the baryon, which encodes the distribution of longitudinal momentum and polarization carried by the quark in the baryon, is defined as
\begin{align}\label{defi_pdf}
\Phi^{\Gamma(q)}(x)=&\frac{1}{2}\int \frac{d z^-}{4\pi} e^{ip^+z^-/2}\nonumber\\
&\times \langle P,\Lambda|\bar{\psi}_{q}(0)\Gamma\psi_{q}(z^-)|P, \Lambda \rangle \bigg|_{z^+=\vec{z}_\perp=0}.
\end{align}
For different Dirac structures, one obtains different quark PDFs of the baryon. For example, for $\Gamma=\gamma^+, \gamma^+ \gamma^5, i\sigma^{j+}\gamma^5$, one has the unpolarized PDF $f(x)$, helicity distribution $g_1(x)$ and transversity distribution $h_1(x)$, respectively. It is worth noting that we work in the light-front gauge $A^+ = 0$, so that the gauge link appearing in between the quark fields in Eq.~(\ref{defi_pdf}) is unity. Here, we compute the quark unpolarized PDFs from the eigenstates of our light-front effective Hamiltonian, Eq.~(\ref{hamilton}), in the constituent valence quarks representation suitable for low-momentum scale applications.

Following the two-point quark correlation function defined in Eq.~(\ref{defi_pdf}), in the LFWFs overlap representation, the
unpolarized PDFs, $f(x)$, in the valence Fock component at the initial scale ($\mu_0$) reads
\begin{align}
f^q(x)=&\sum_{\lambda_i} \int \left[{\rm d}\mathcal{X} \,{\rm d}\mathcal{P}_\perp\right]\nonumber\\ &\times\Psi^{\uparrow *}_{\{x_i,\vec{p}_{i\perp},\lambda_i\}}\Psi^{\uparrow}_{\{x_i,\vec{p}_{i\perp},\lambda_i\}} \delta \left(x-x_q\right).
\end{align}
Using the LFWFs within the BLFQ approach given in Eq.~(\ref{wavefunctions}), we evaluate the unpolarized PDFs for the valence quarks in the baryon, which are normalized as
\begin{align}
&\int_{0}^{1} dxf^q(x)=F^{\rm{q}}_1(0)=n_q,
\end{align}
with $n_q$ being the number of quarks of flavor $q$ in the baryon.
Furthermore, at the model scale, the following momentum sum rule is satisfied by our PDFs,
\begin{align}
\sum_q\int_{0}^{1} dx\, xf^q(x) = 1. 
\end{align}

By performing the QCD evolution, we obtain the valence quark PDFs at higher $\mu^2$ scales using input PDFs at the model scale. 
We interpret the model scale associated with our LFWFs as the effective scale where the structures of the baryon are described by the motion of the valence quarks only.
The scale evolution allows the valence quarks to produce gluons, with the emitted gluons capable of producing quark-antiquark pairs as well as additional gluons. In this picture, the gluon and sea quark components emerge at scales higher than the model scale.

The QCD evolution is described by the well-known Dokshitzer-Gribov-Lipatov-Altarelli-Parisi (DGLAP) equations~\cite{Dokshitzer:1977sg,Gribov:1972ri,Altarelli:1977zs}. Here, we utilize the next-to-next-to-leading order (NNLO) DGLAP equations of QCD, to evolve our valence quark distributions from our model scale $\mu_0^2$ to higher scale $\mu^2$. 
For this purpose, we employ the Higher Order Perturbative Parton Evolution toolkit (HOPPET) to numerically solve the NNLO DGLAP equations~\cite{Salam:2008qg}.
We adopt $\mu_0^2=0.45\pm 0.04$ GeV$^2$ for the initial scale of our PDFs for $\Lambda(\Sigma^0,\,\Sigma^+,\,\Sigma^-)$, which we determine by matching the moment of the valence quark PDFs for $\Lambda$: $\langle{x}\rangle_q=\int_0^1 \,{\rm d}x\, x\, f_1^q(x)$ at $\mu^2=4$ GeV$^2$, with the result from the lattice QCD simulations $\langle{x}\rangle_u=\langle{x}\rangle_d=0.20\pm 0.01$ and $\langle{x}\rangle_s=0.27\pm0.01$~\cite{Gockeler:2002uh}, after performing the QCD evolution of our initial quark PDFs. Meanwhile, we choose the initial scale for the PDFs of $\Lambda_c(\Sigma^+_c,\,\Sigma^{++}_c,\,\Sigma^0_c)$ as the hadronic scale $\mu_0^2=1.0\pm 0.1$ GeV$^2$.

\begin{figure*}[tph]
	\includegraphics[width=0.35\textwidth]{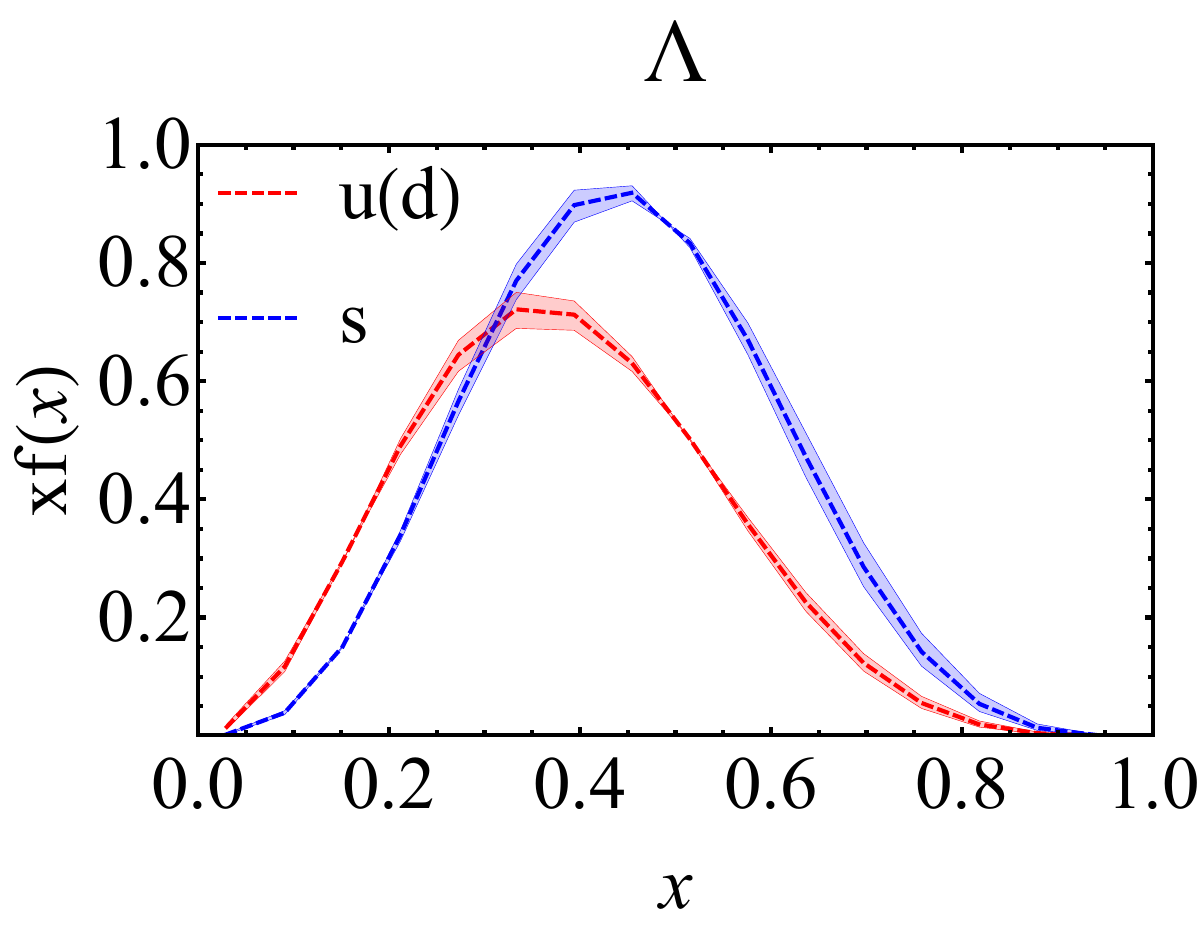}
	\includegraphics[width=0.35\textwidth]{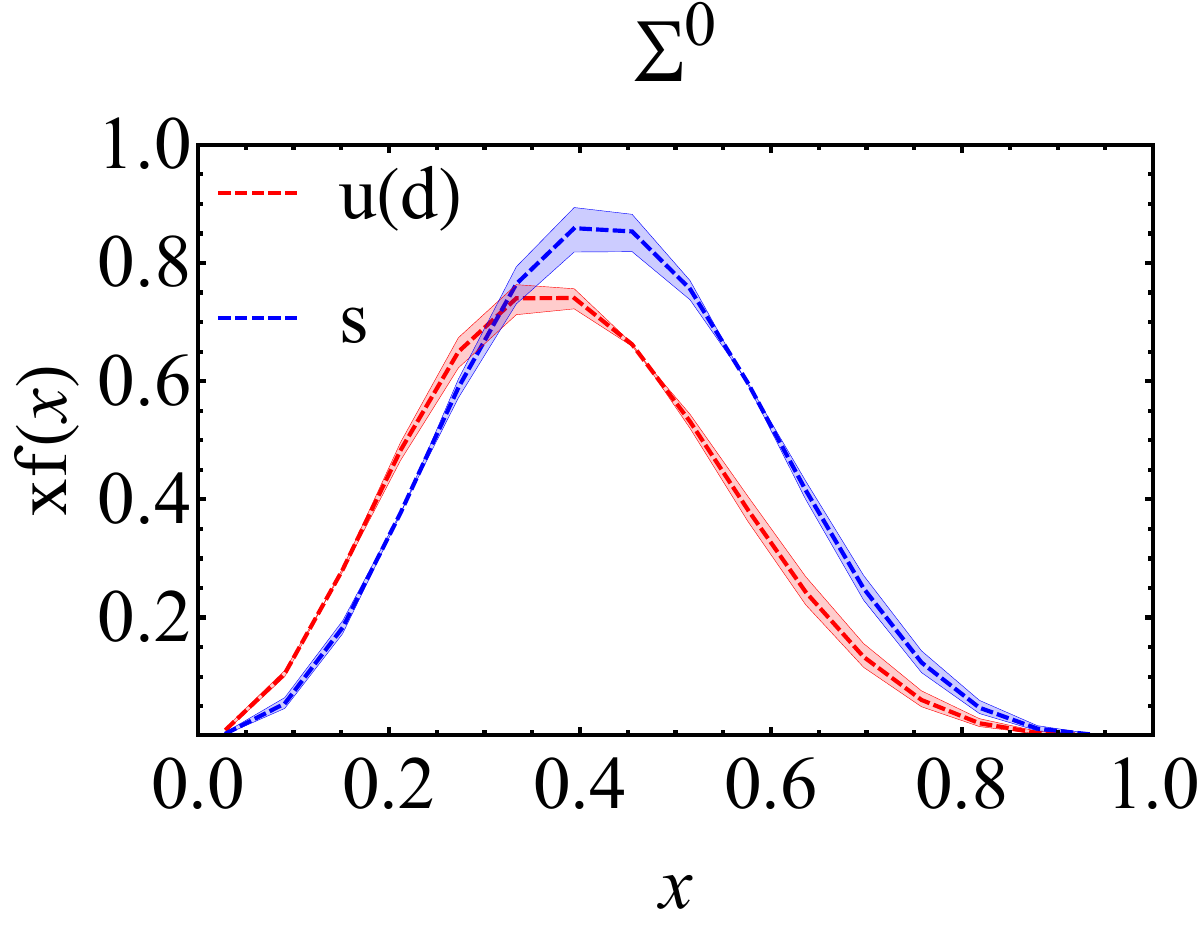}
	\includegraphics[width=0.35\textwidth]{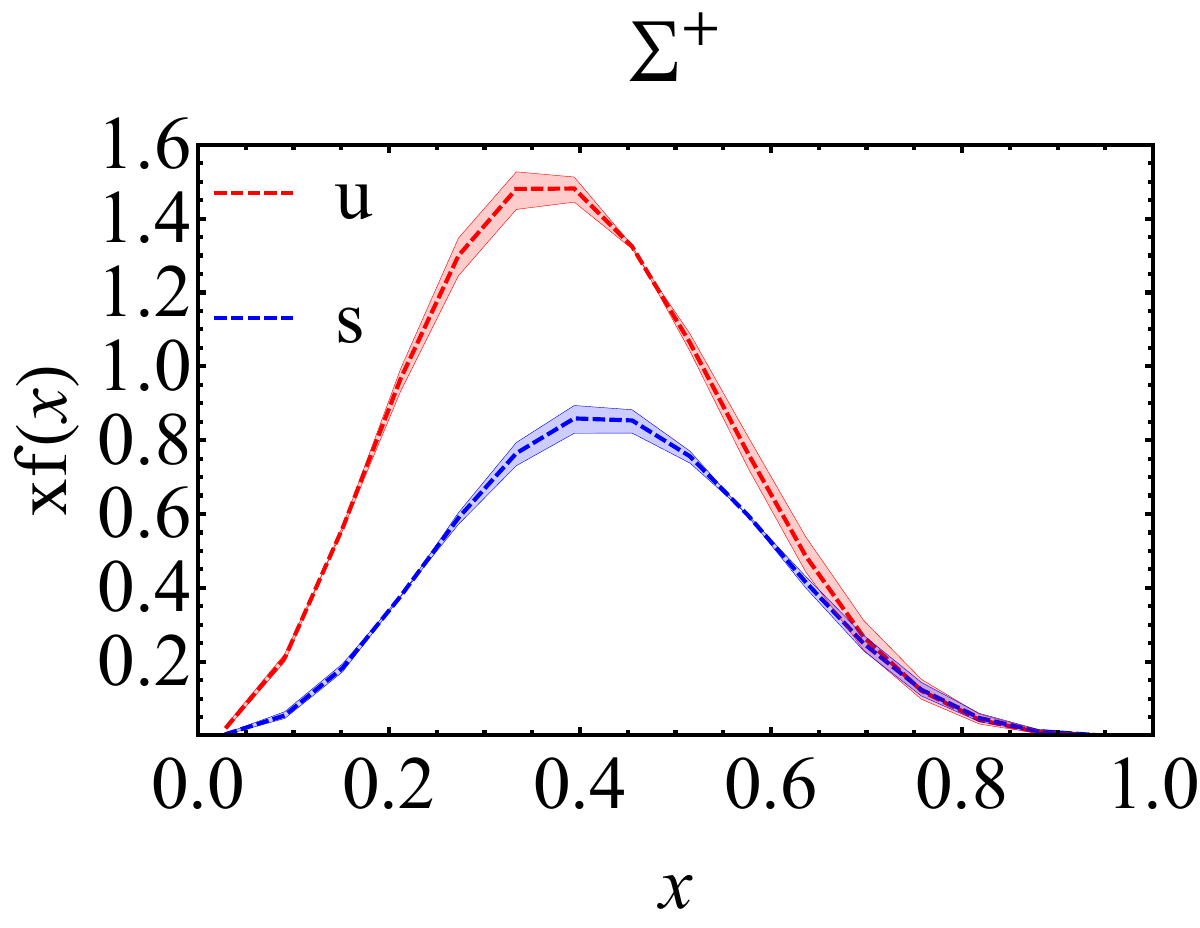}
	\includegraphics[width=0.35\textwidth]{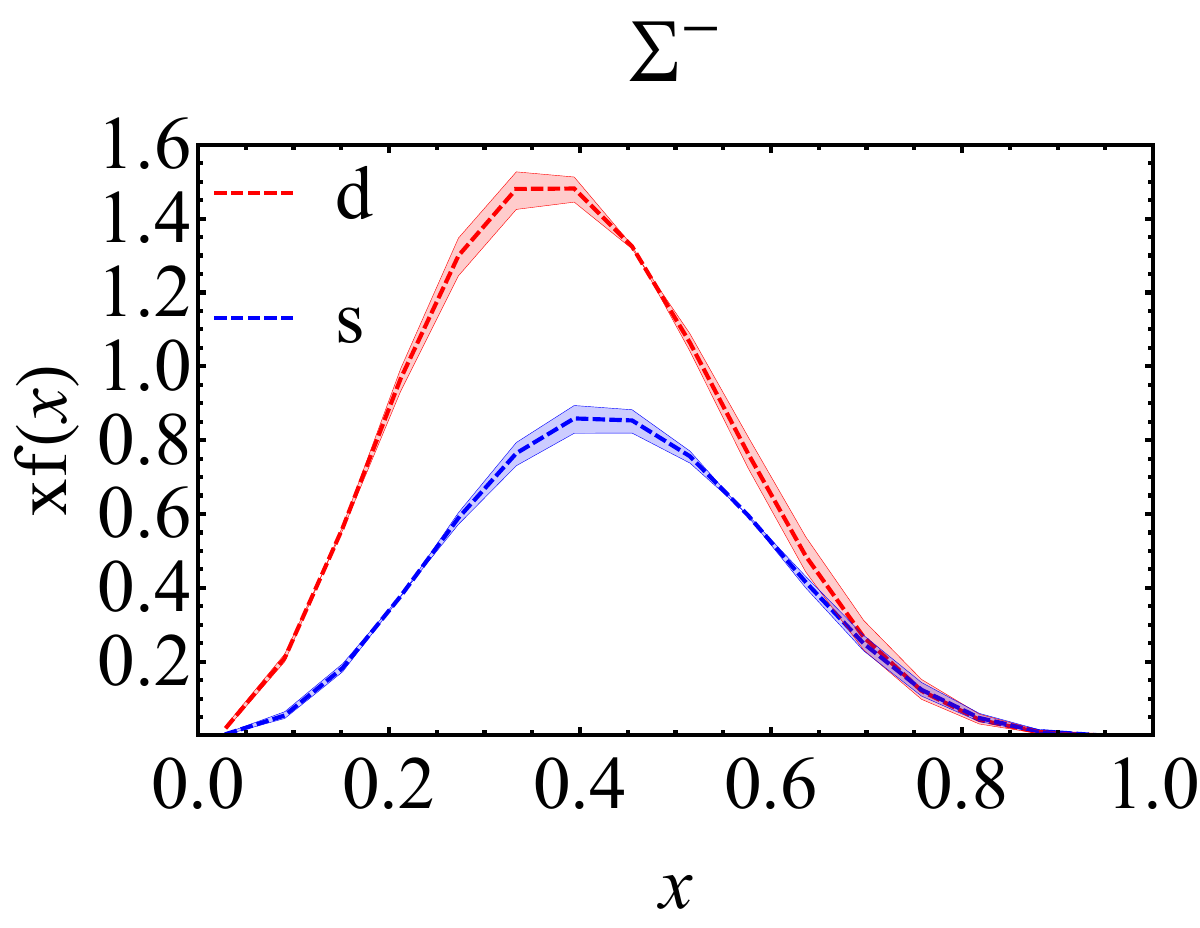}
	\caption{Valence quarks' unpolarized PDFs at the model scale  multiplied by $x$ as functions of $x$. Upper-left panel: $\Lambda$; upper-right panel: $\Sigma^0$; lower-left panel: $\Sigma^+$; lower-right panel: $\Sigma^-$.  The blue and red bands represent distributions for the strange quark and the light quark ($u/d$), respectively. The bands reflect the $10\%$ uncertainty  in the coupling constant $\alpha_s$.}
	\label{figs:lambda-xpdf}
\end{figure*}
\begin{figure*}[tph]
		\includegraphics[width=0.35\textwidth]{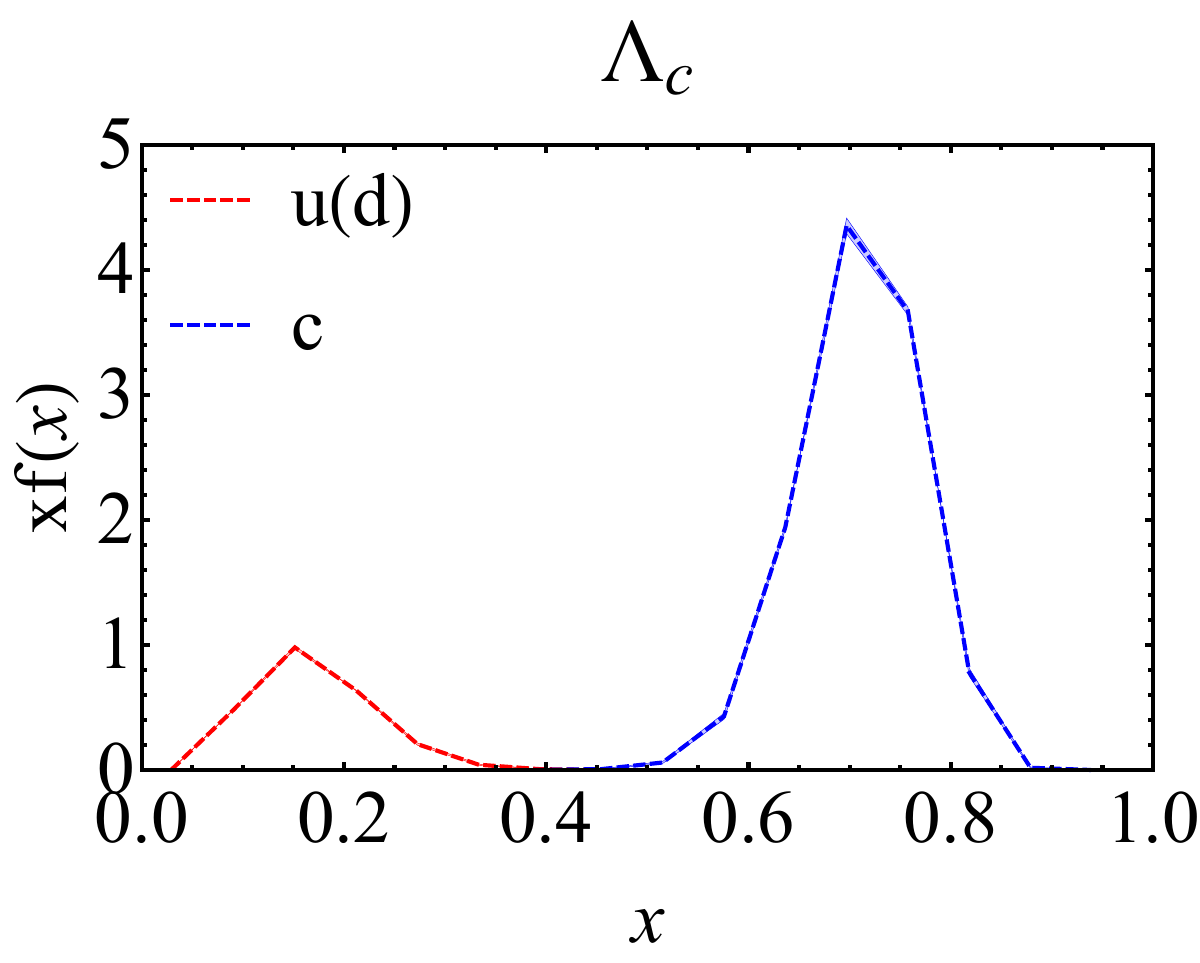}
	\includegraphics[width=0.35\textwidth]{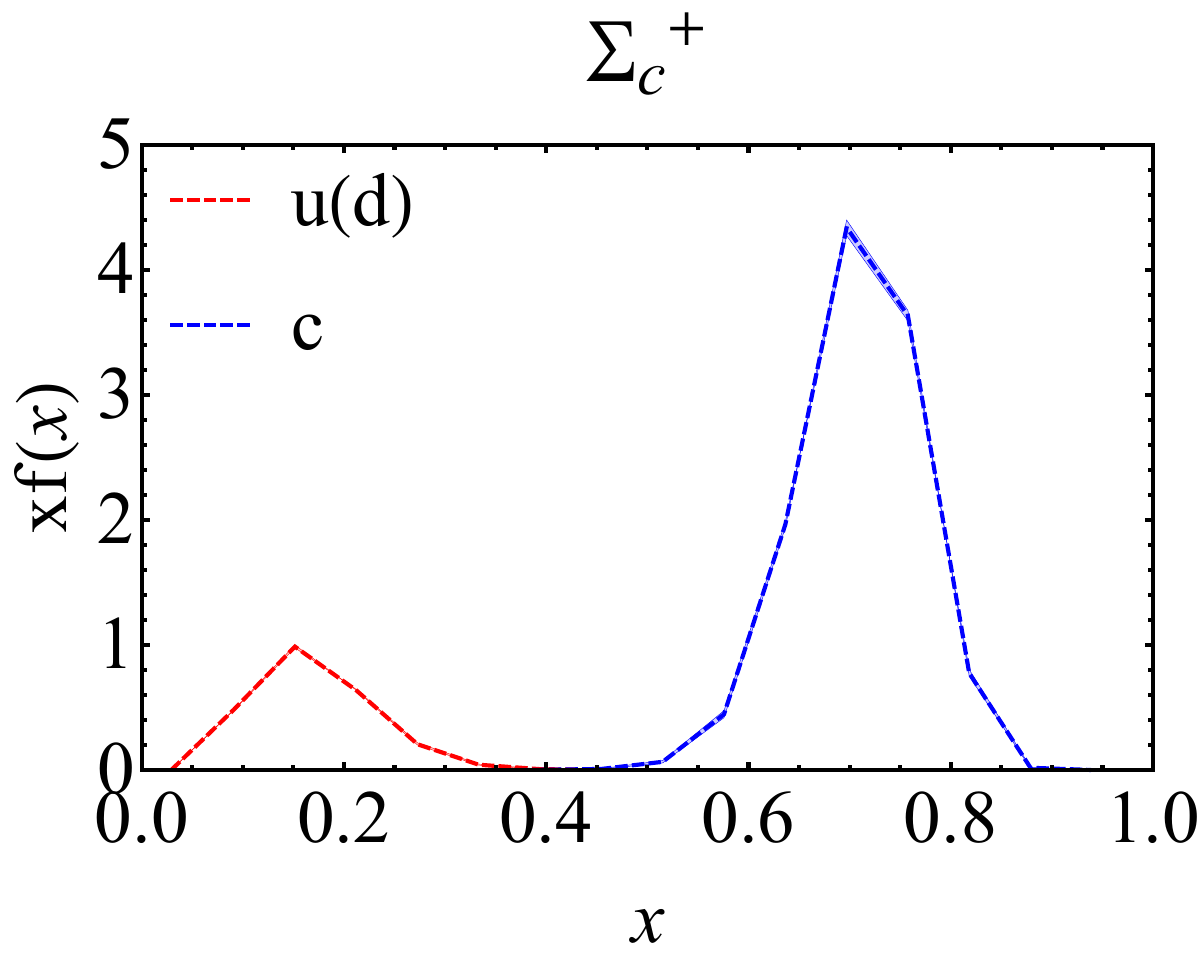}
	\includegraphics[width=0.35\textwidth]{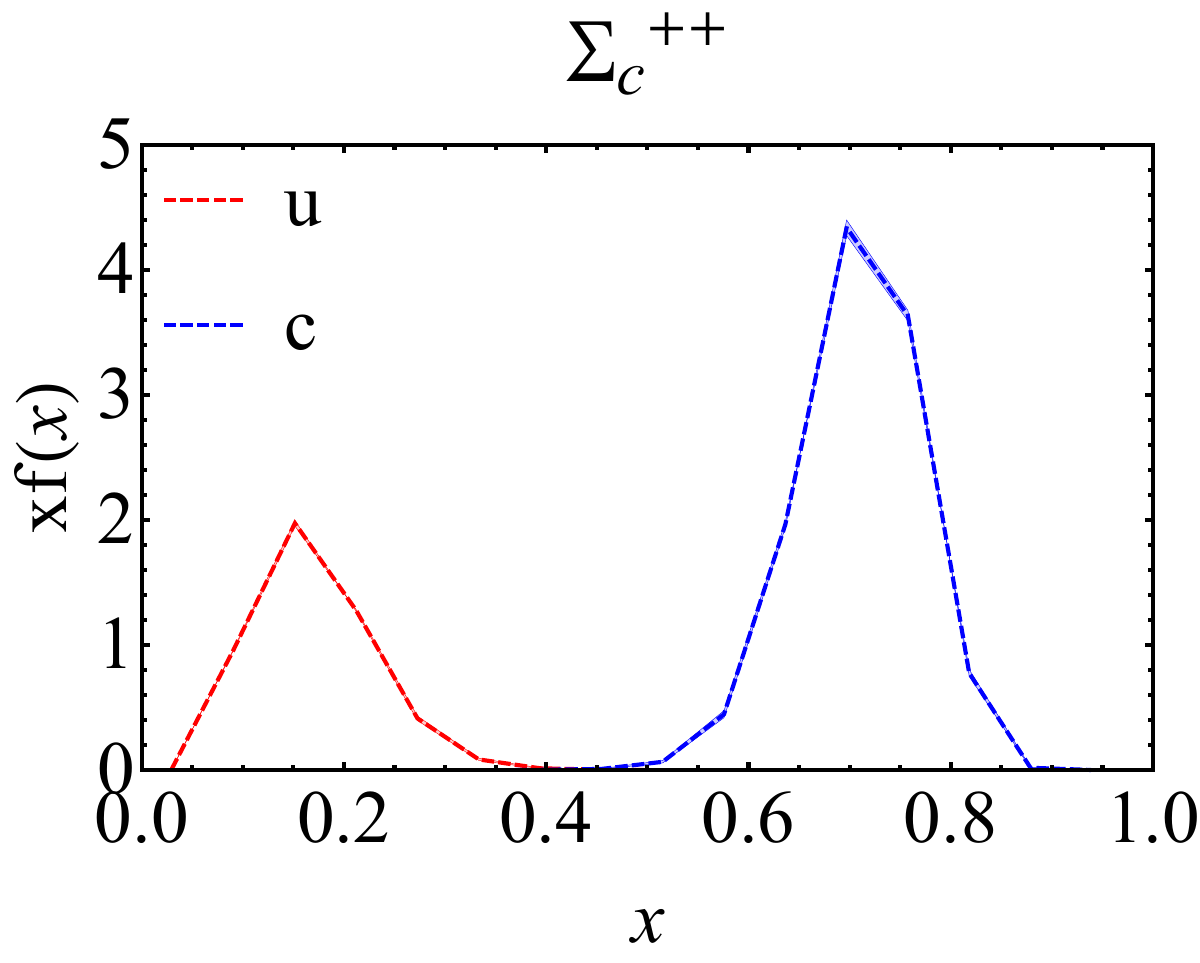}
	\includegraphics[width=0.35\textwidth]{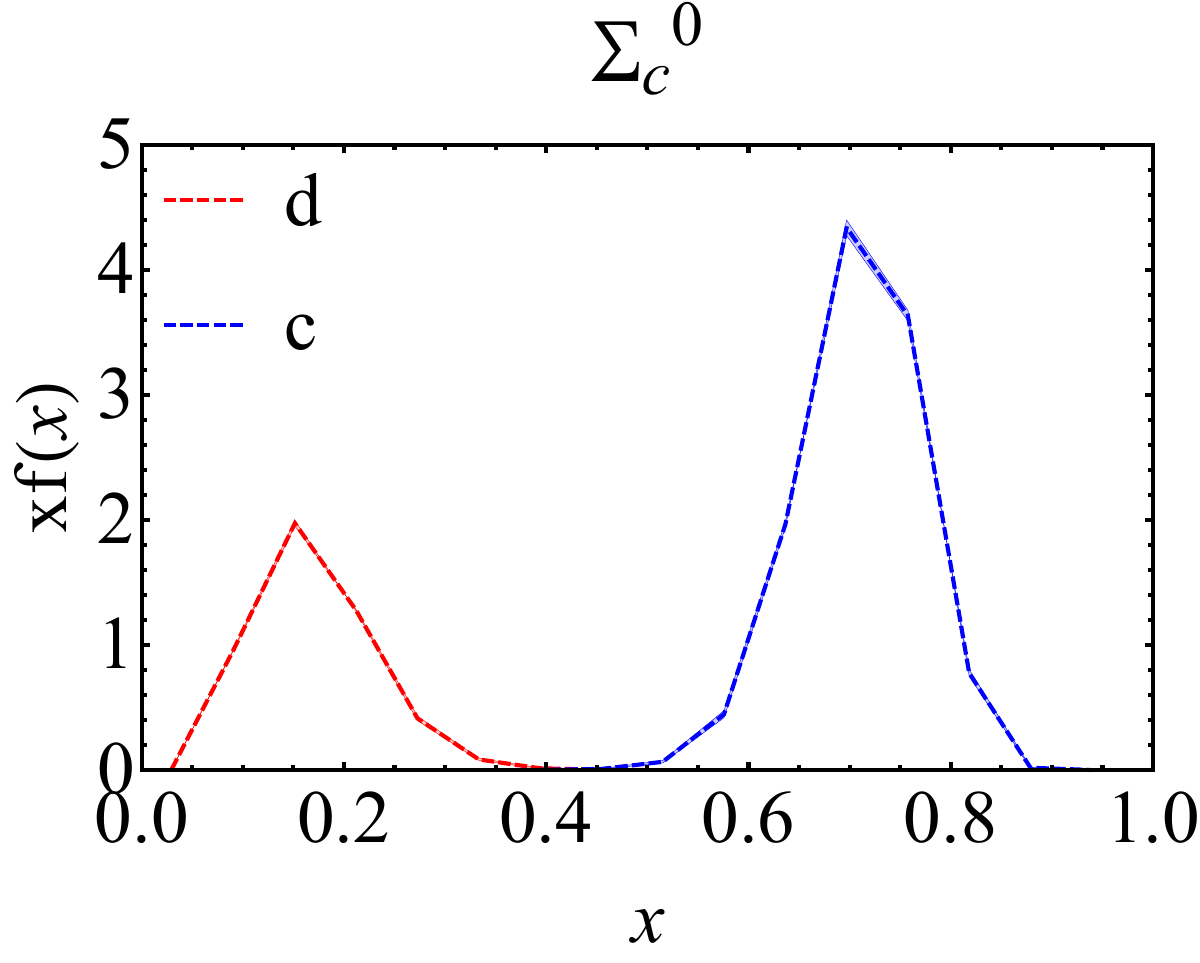}
	\caption{Valence quarks' unpolarized PDFs at the model scale  multiplied by $x$ as functions of $x$. Upper-left panel: $\Lambda_c$; upper-right panel: $\Sigma_c^+$; lower-left panel: $\Sigma_c^{++}$; lower-right panel: $\Sigma_c^{0}$.  The blue and red bands represent distributions for the charm quark and the light quark ($u/d$), respectively. The bands reflect the $10\%$ uncertainty in the coupling constant $\alpha_s$.}
	\label{figs:lambdac-xpdf}
\end{figure*}
\begin{figure*}[tph]
	\includegraphics[width=0.35\textwidth]{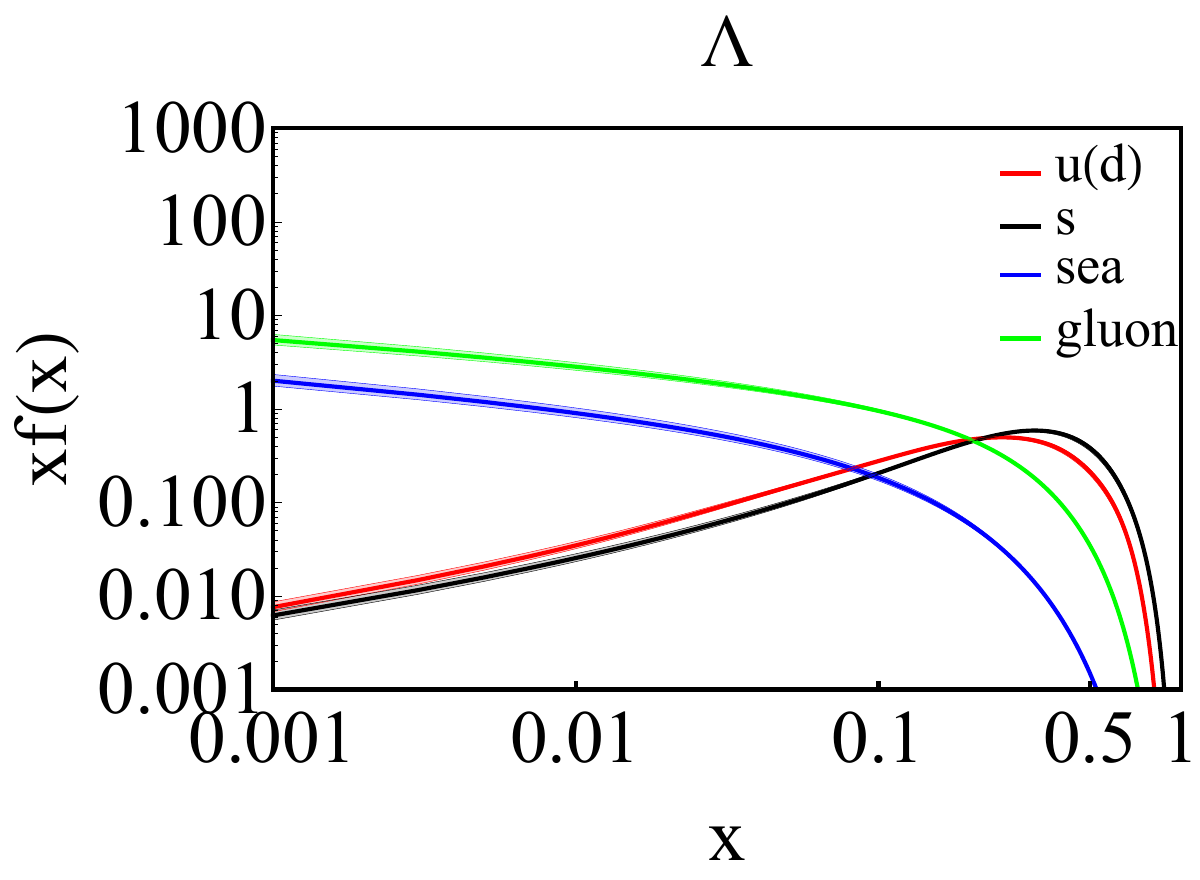}
	\includegraphics[width=0.35\textwidth]{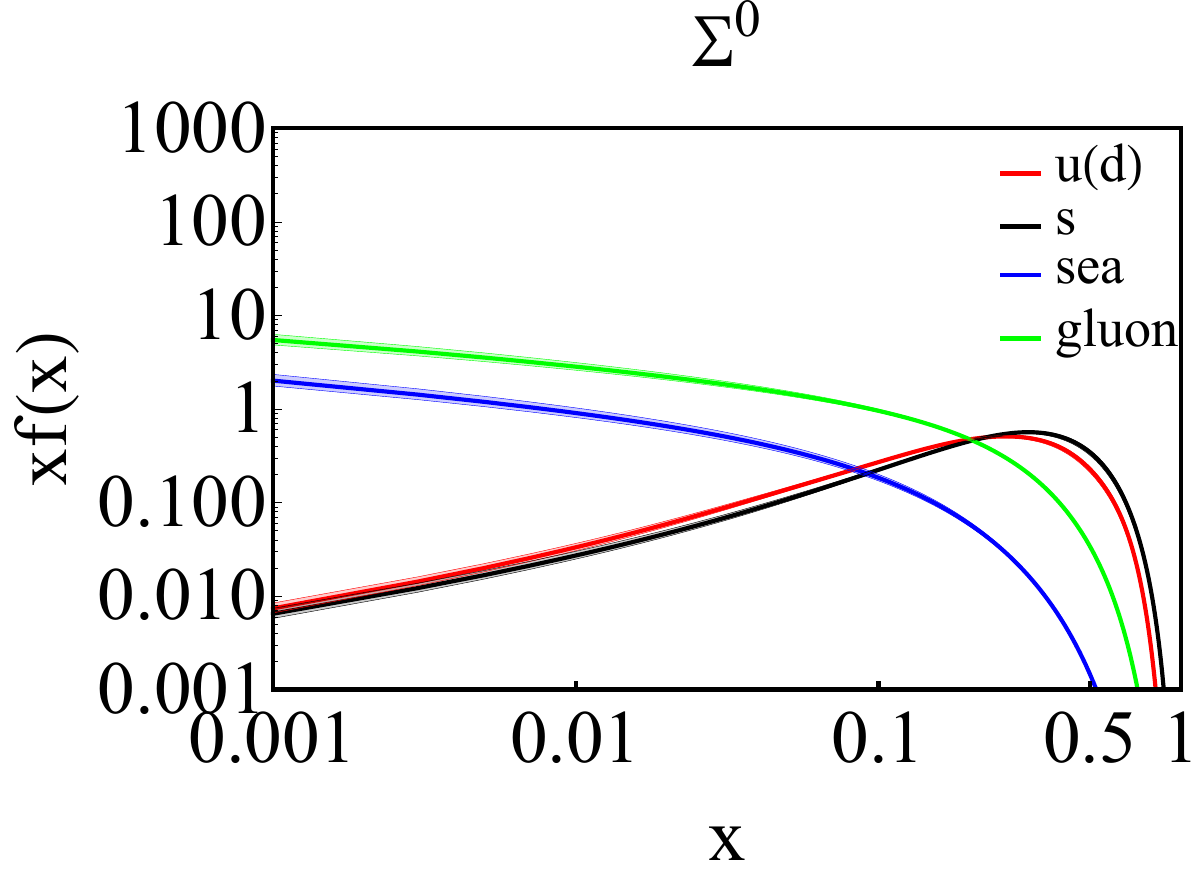}
	\includegraphics[width=0.35\textwidth]{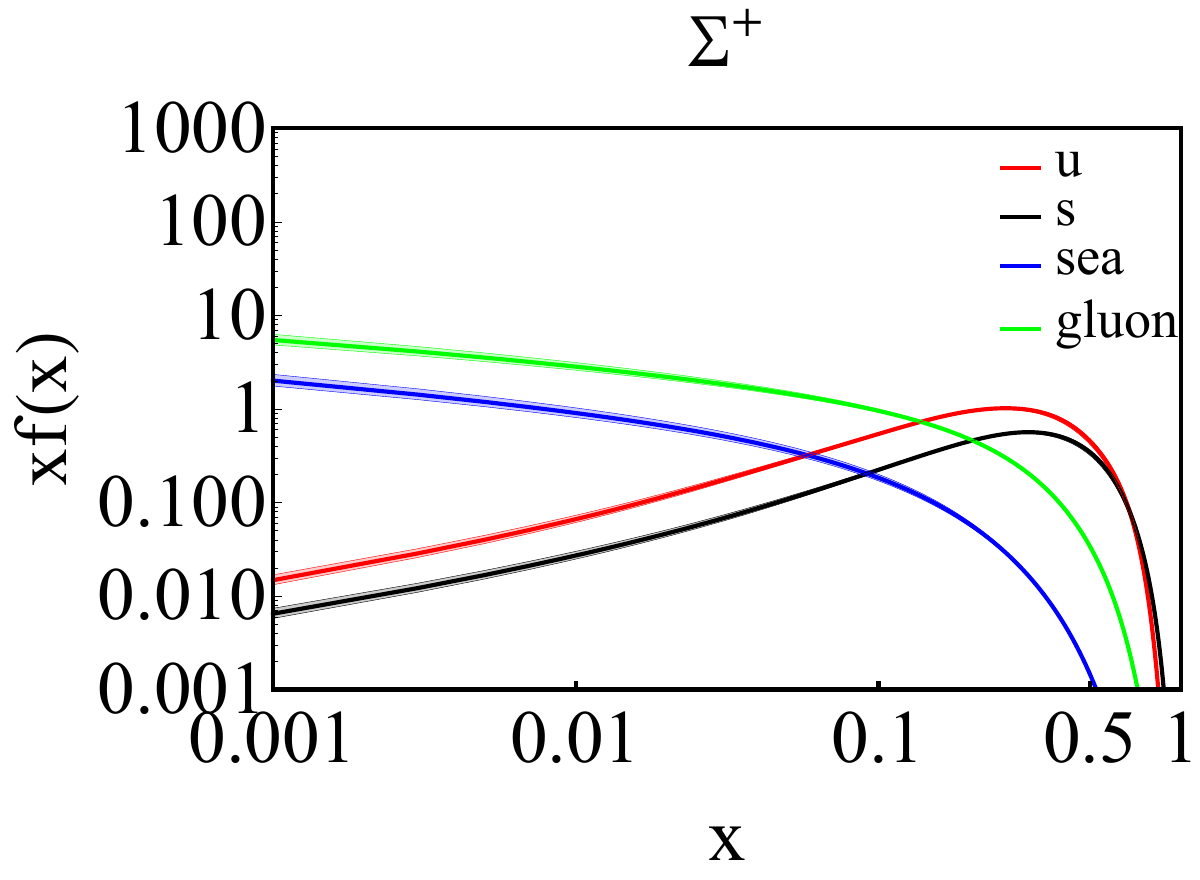}
	\includegraphics[width=0.35\textwidth]{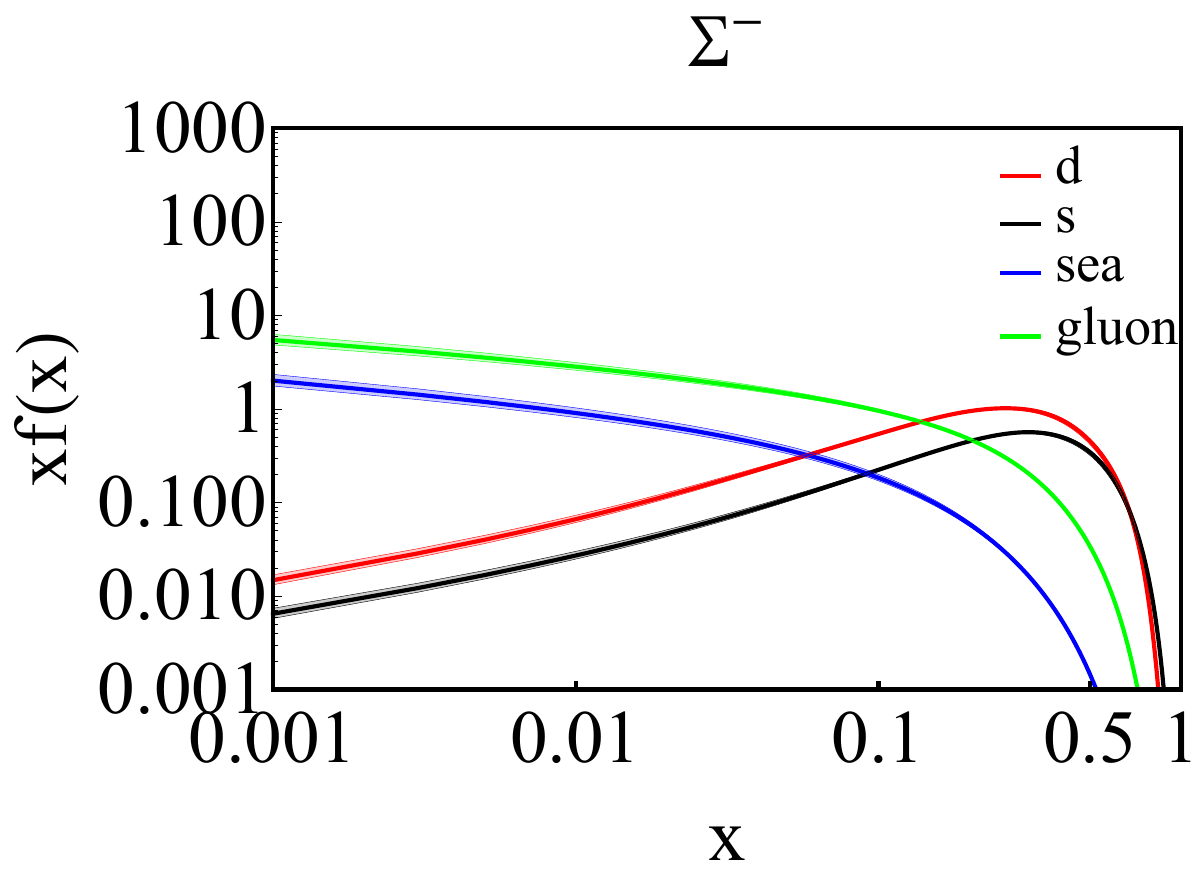}
		\caption{Unpolarized PDFs at $10$ GeV$^2$  multiplied by $x$ as functions of $x$. Upper-left panel: $\Lambda$; upper-right panel: $\Sigma^0$; lower-left panel: $\Sigma^+$; lower-right panel: $\Sigma^-$.  The red, black, blue, and green bands represent distributions for the valence light quark ($u/d$), the valence strange quark, sea quarks, and gluon, respectively. The bands reflect the $10\%$ uncertainty in the initial scale $\mu_0$.}
	\label{fig:lambda-10GeV2}
\end{figure*}
\begin{figure*}[tph]
		\includegraphics[width=0.35\textwidth]{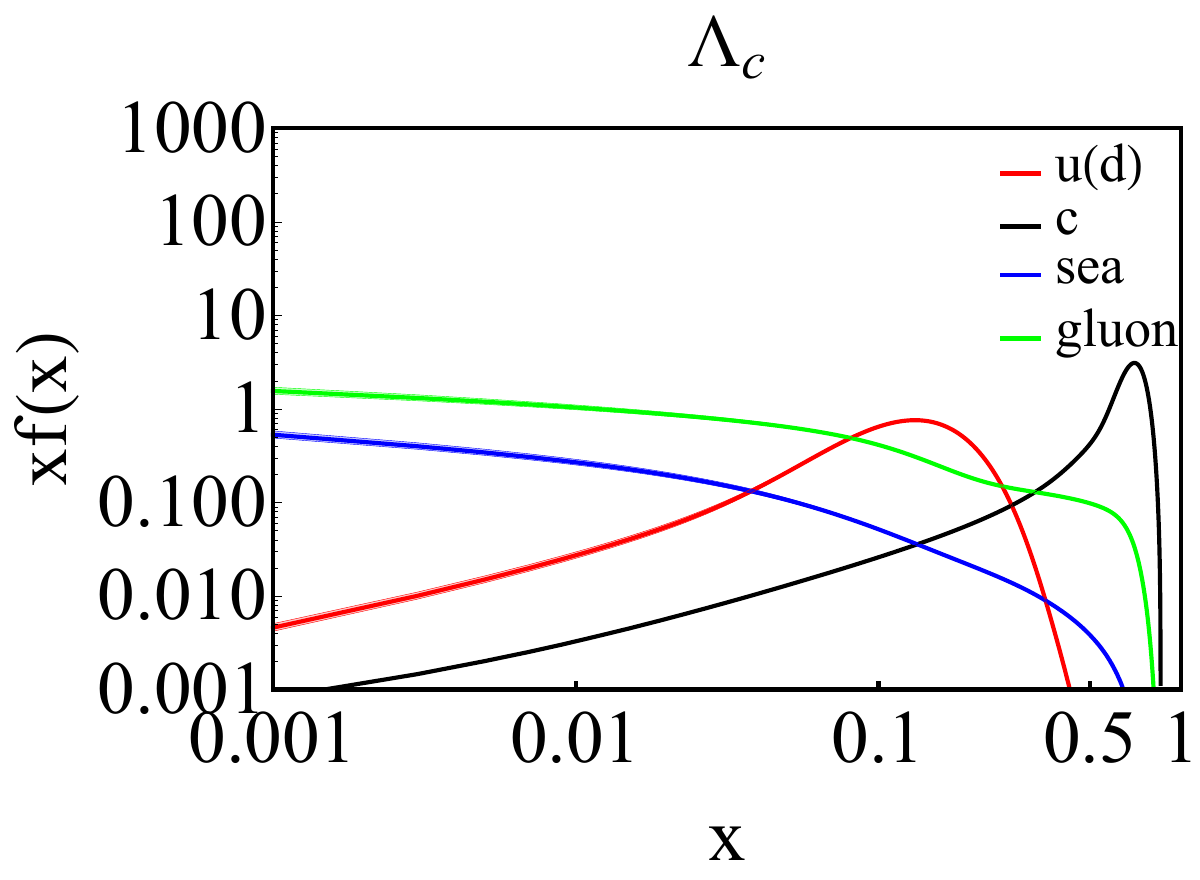}
	\includegraphics[width=0.35\textwidth]{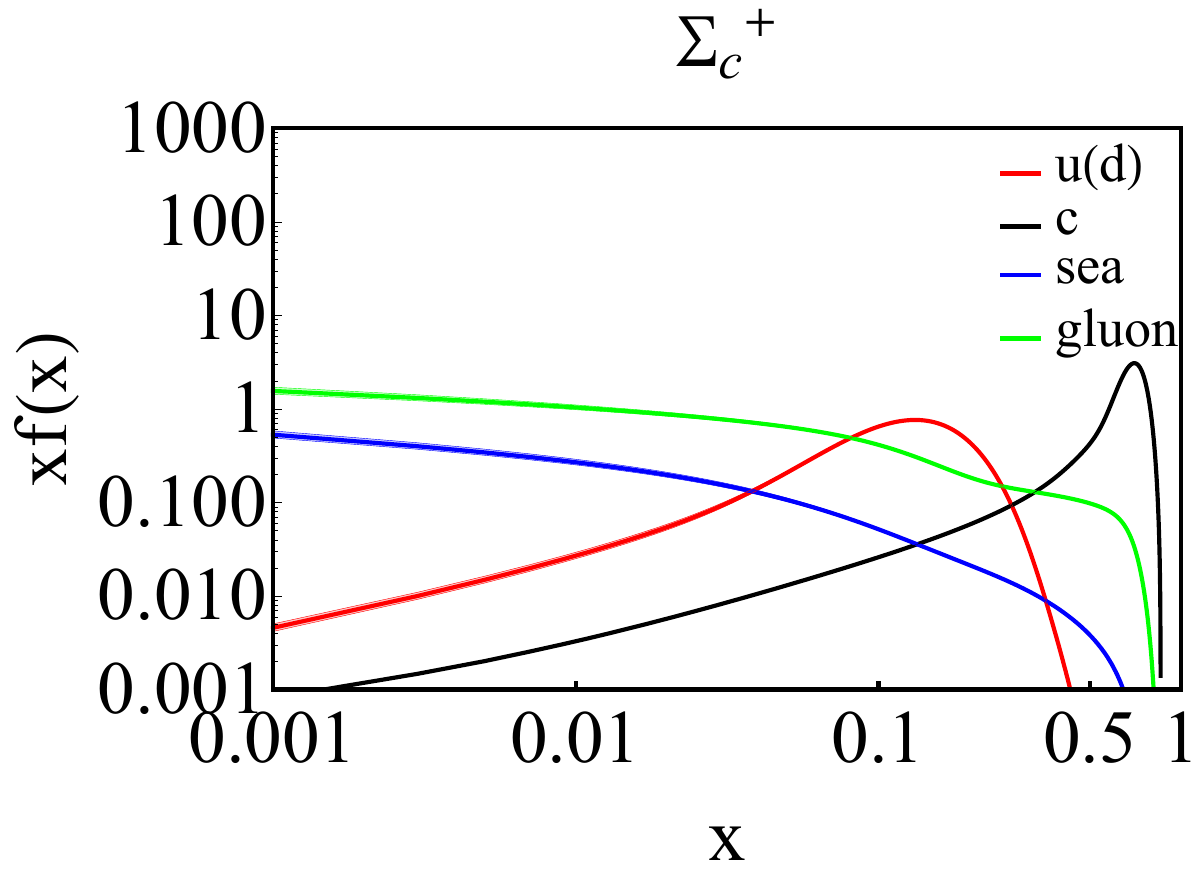}
	\includegraphics[width=0.35\textwidth]{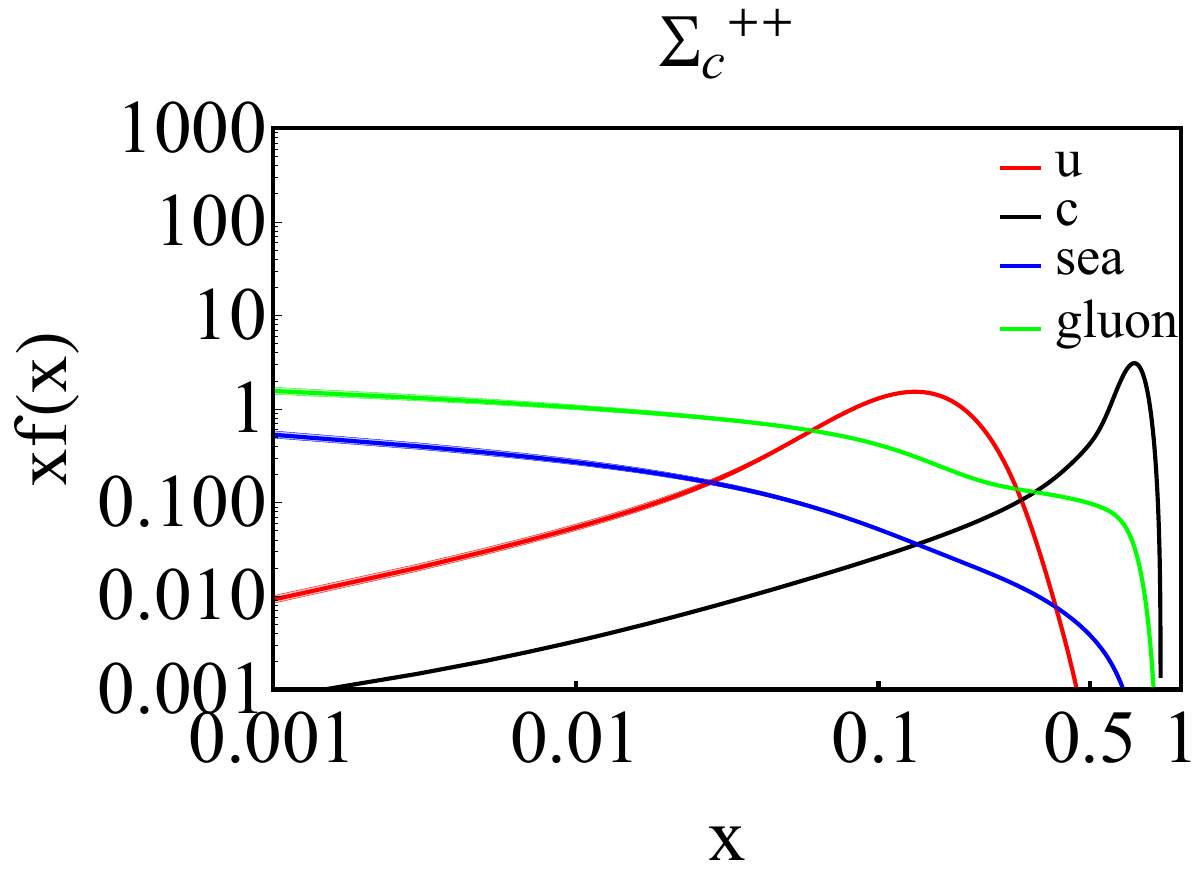}
	\includegraphics[width=0.35\textwidth]{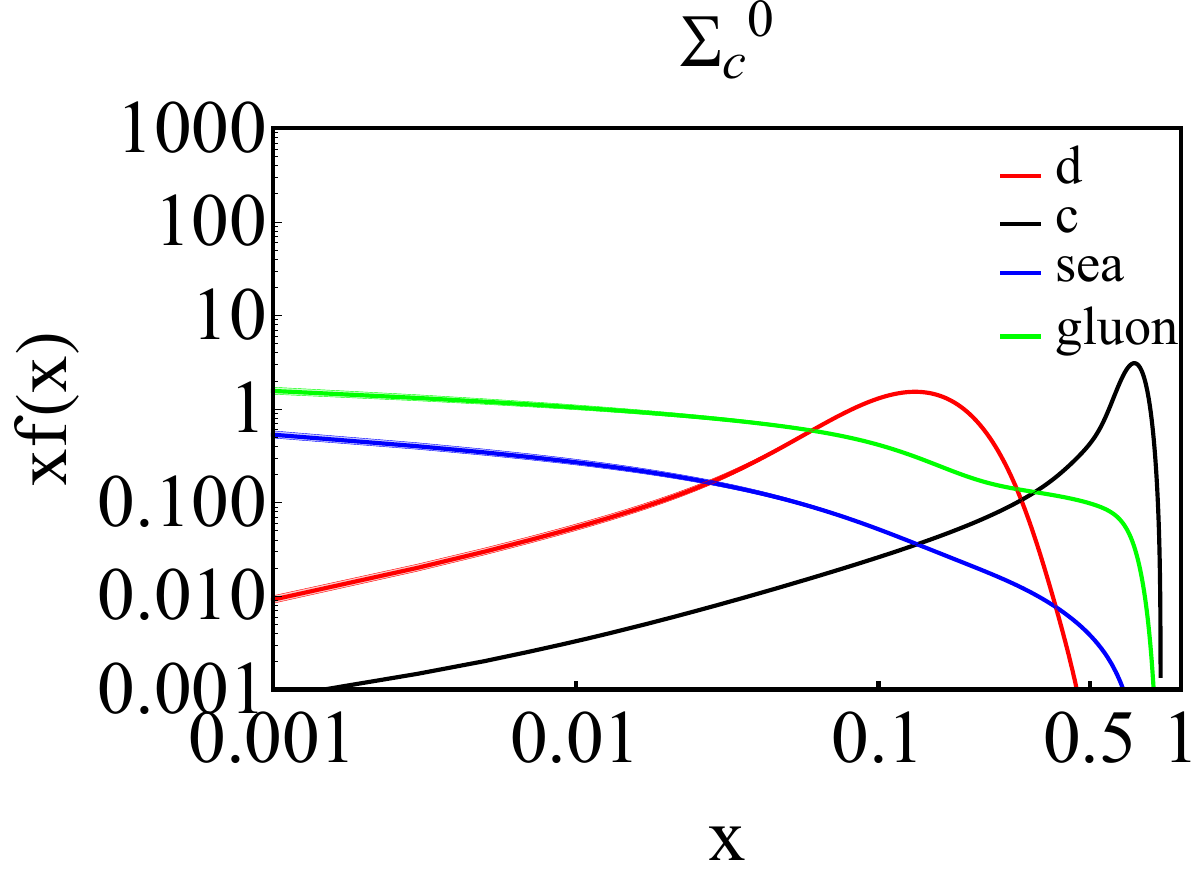}
		\caption{Unpolarized PDFs at $10$ GeV$^2$  multiplied by $x$ as functions of $x$. Upper-left panel: $\Lambda_c$; upper-right panel: $\Sigma_c^+$; lower-left panel: $\Sigma_c^{++}$; lower-right panel: $\Sigma_c^{0}$.  The red, black, blue, and green bands represent distributions for the valence light quark ($u/d$), the valence charm quark, sea quarks, and gluon, respectively. The bands reflect the $10\%$ uncertainty in the initial scale $\mu_0$.}
	\label{fig:lambdac-10GeV2}
\end{figure*}
\begin{figure*}[tph]
	\includegraphics[width=0.35\textwidth]{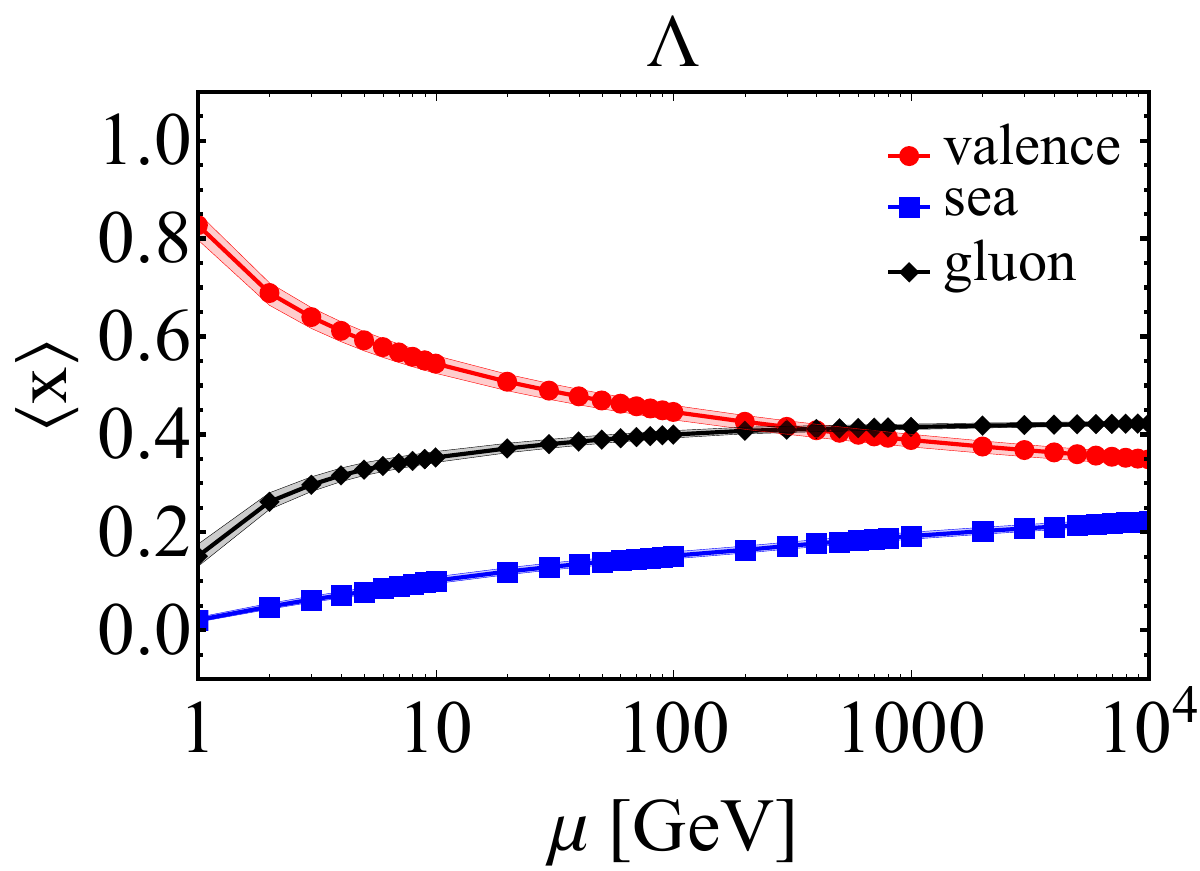}
	\includegraphics[width=0.35\textwidth]{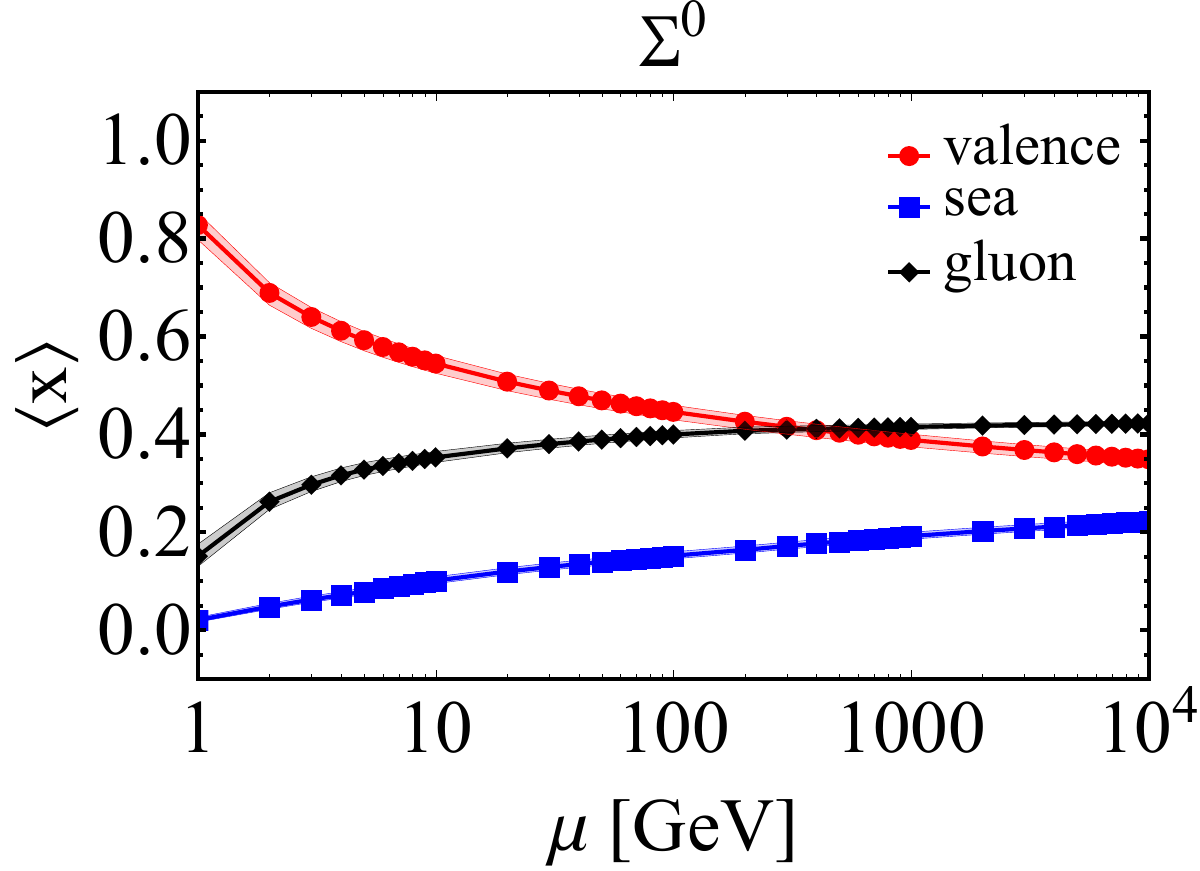}
	\includegraphics[width=0.35\textwidth]{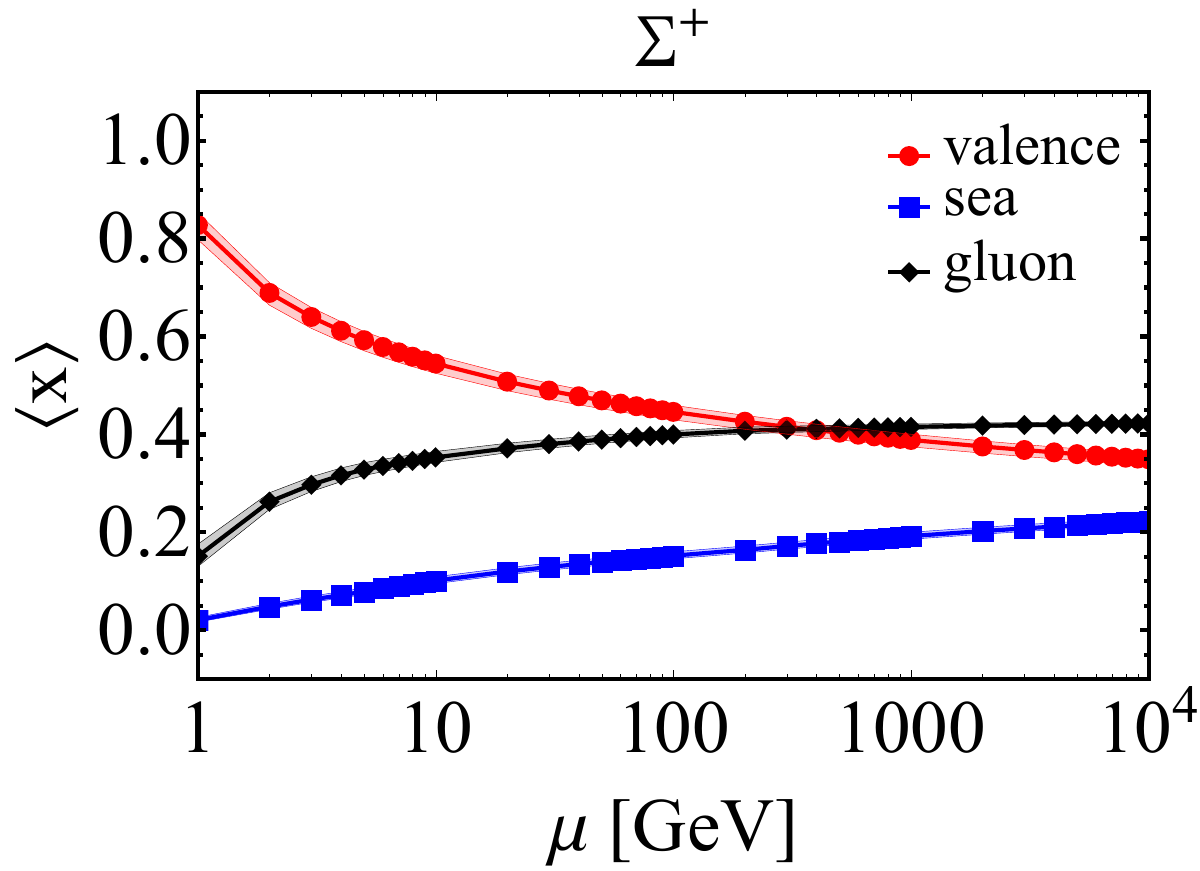}
	\includegraphics[width=0.35\textwidth]{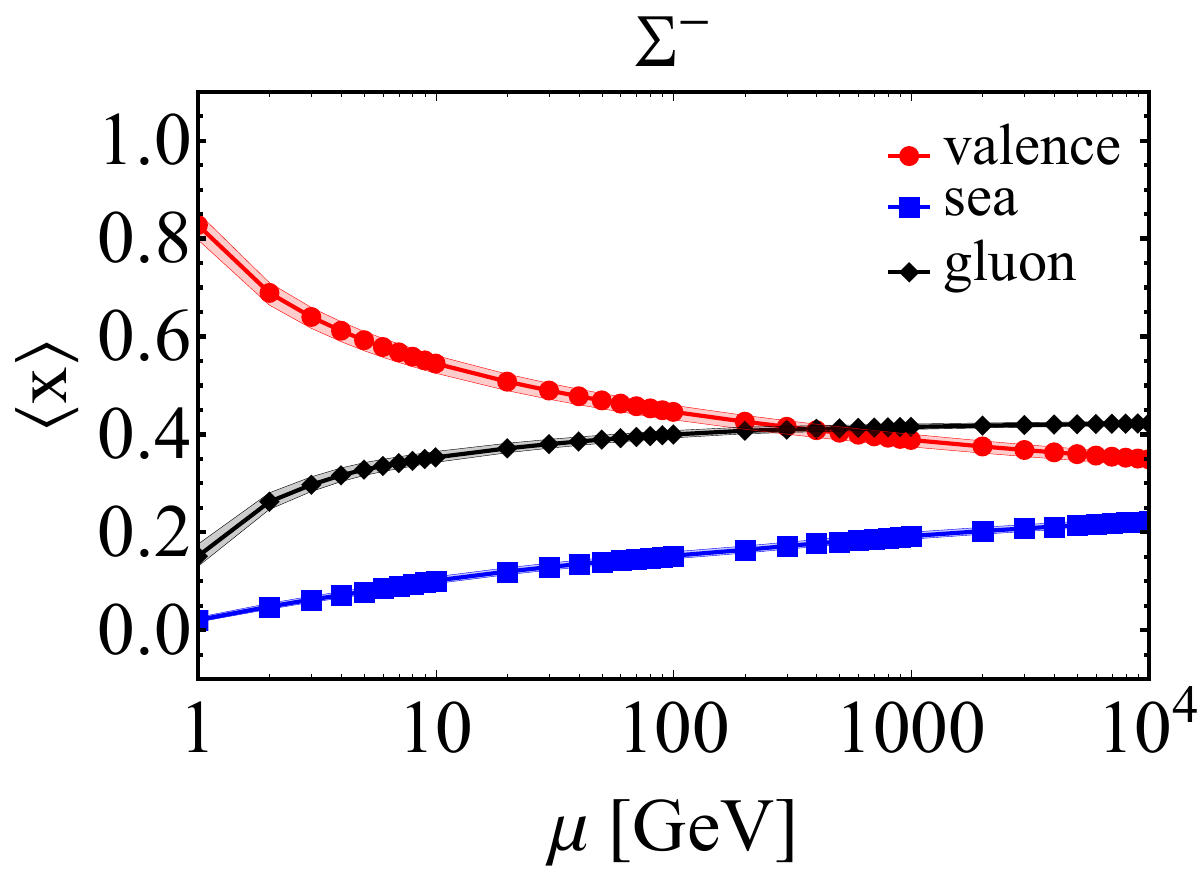}	
	\caption{The first moment of the PDFs of $\Lambda$ (upper-left panel),  $\Sigma^0$ (upper-right panel), $\Sigma^{+}$ (lower-left panel), and $\Sigma^{-}$ (lower-right panel) as functions of the scale $\mu$. The red, black, and blue bands represent the first moments of valence quark, sea quark, and gluon, respectively.   The bands reflect the $10\%$ uncertainty in the initial scale $\mu_0$. }
	\label{fig:lambda-x1}
\end{figure*}
\begin{figure*}[tph]
	\includegraphics[width=0.35\textwidth]{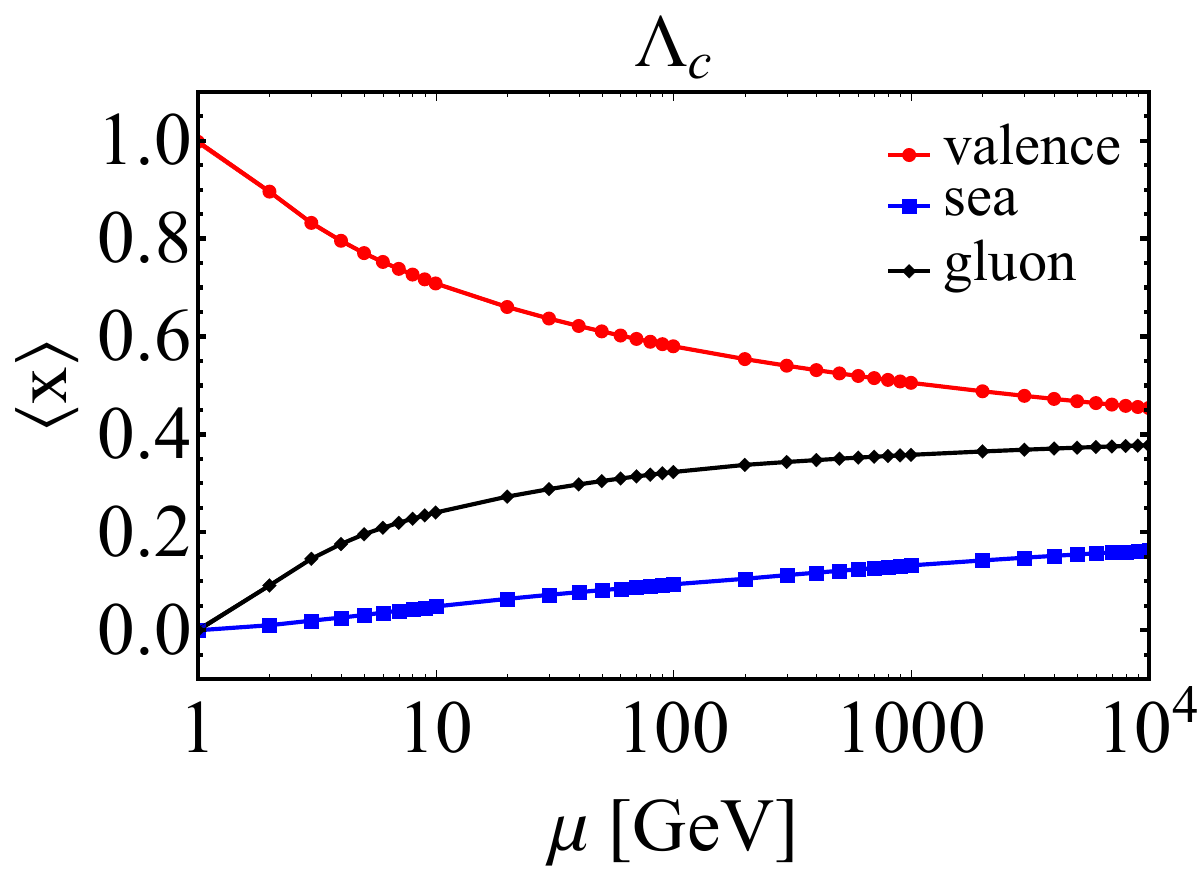}
	\includegraphics[width=0.35\textwidth]{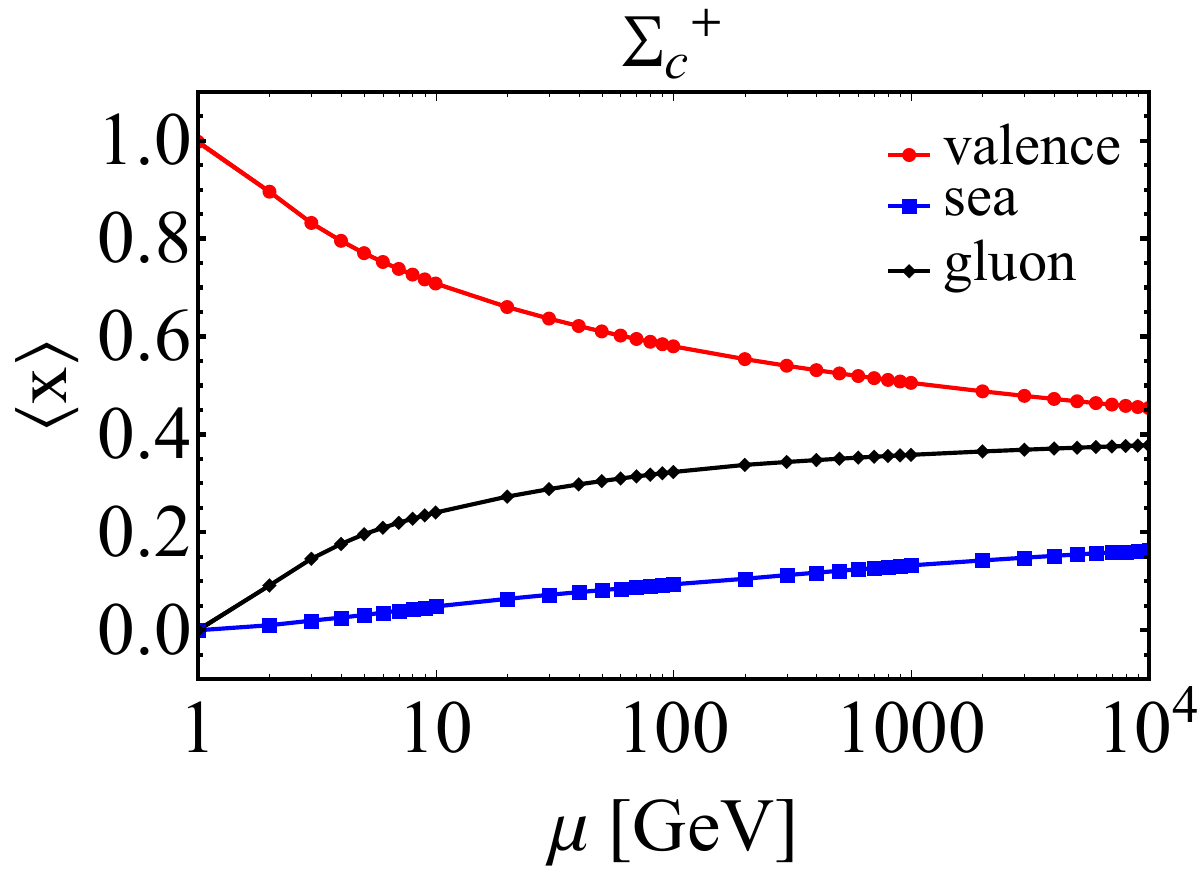}
	\includegraphics[width=0.35\textwidth]{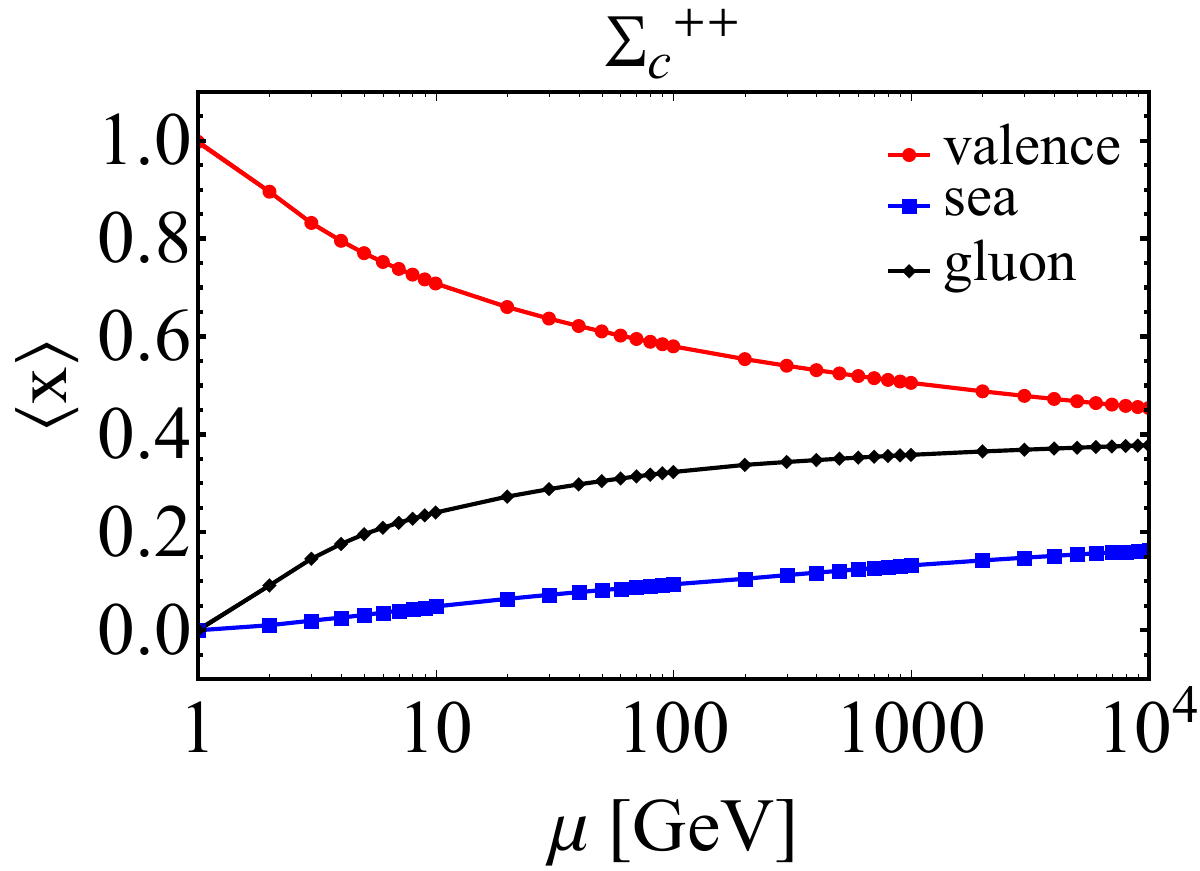}
	\includegraphics[width=0.35\textwidth]{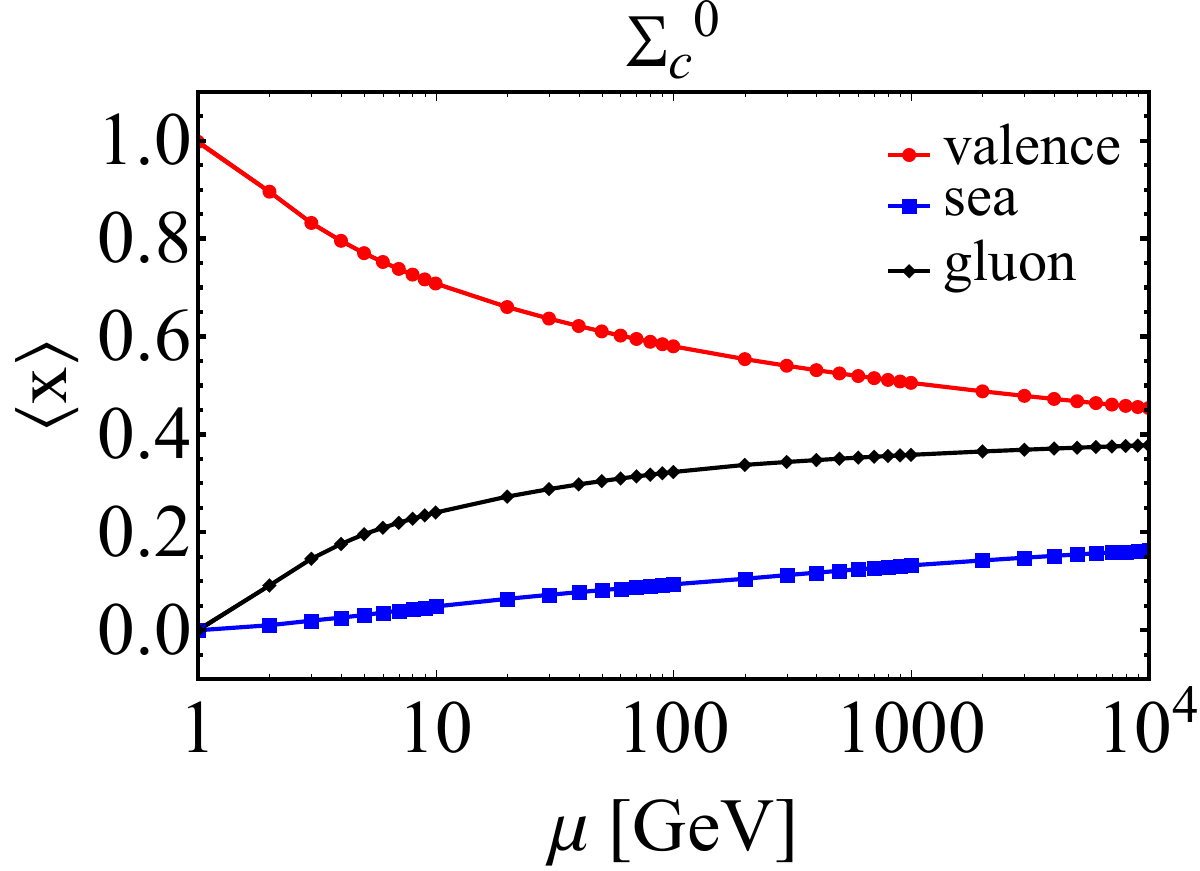}	
	\caption{The first moment of the PDFs of $\Lambda_c$ (upper-left panel),  $\Sigma_c^+$ (upper-right panel), $\Sigma_c^{++}$ (lower-left panel), and $\Sigma_c^{0}$ (lower-right panel) as functions of the scale $\mu$. The red, black, and blue bands represent the first moments of valence quark, sea quark, and gluon, respectively.   The bands reflect the $10\%$ uncertainty in the initial scale $\mu_0$. }
	\label{fig:lambdac-x1}
\end{figure*}
\begin{figure*}[tph]
	\includegraphics[width=0.4\textwidth]{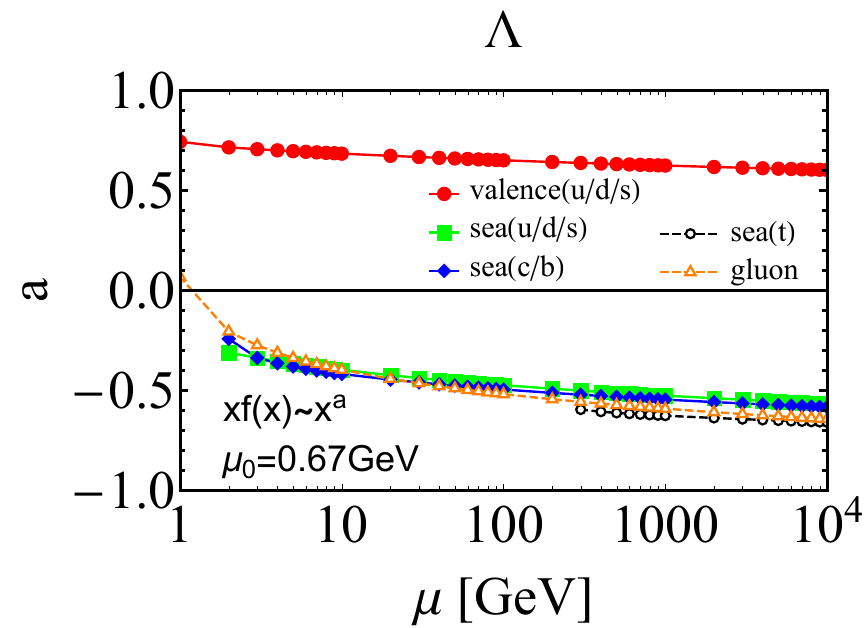}
	\includegraphics[width=0.4\textwidth]{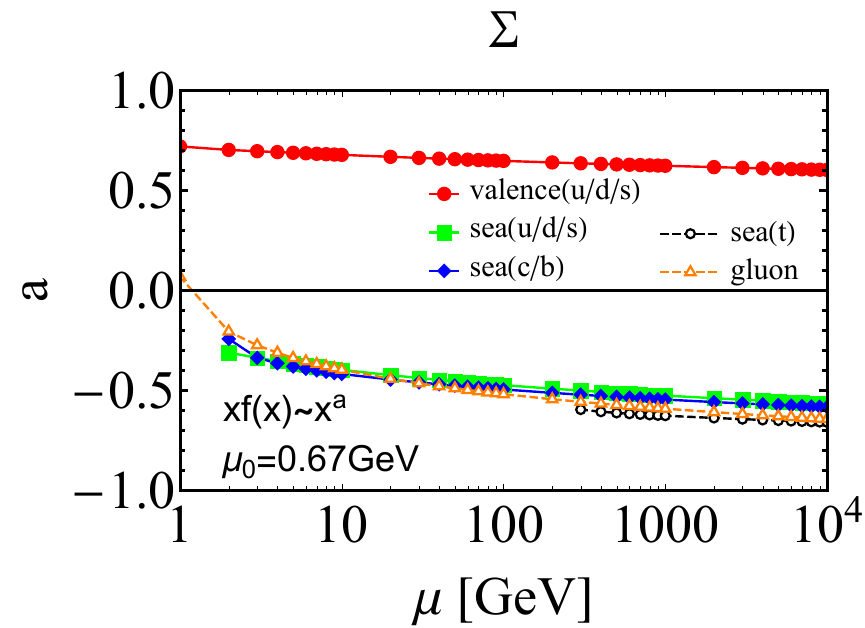}
	\includegraphics[width=0.4\textwidth]{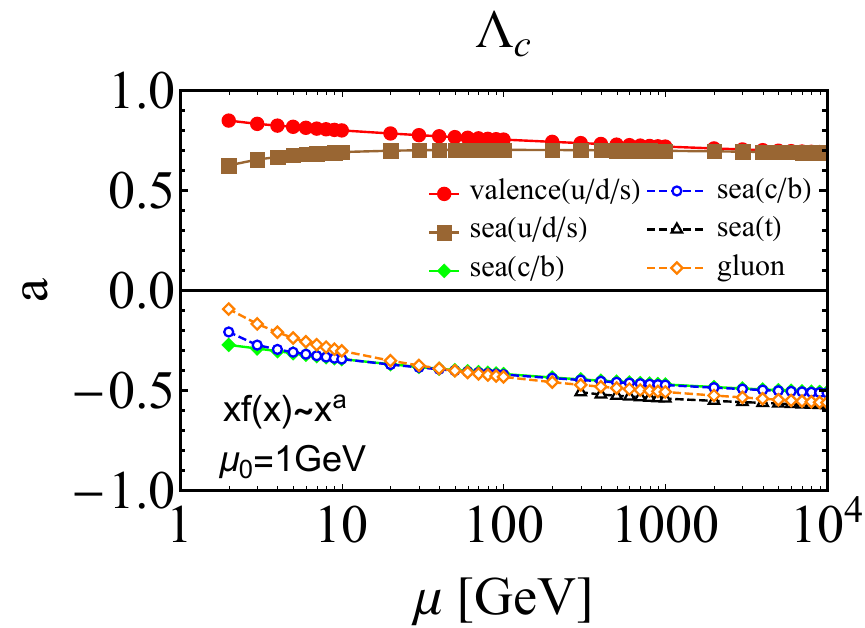}
	\includegraphics[width=0.4\textwidth]{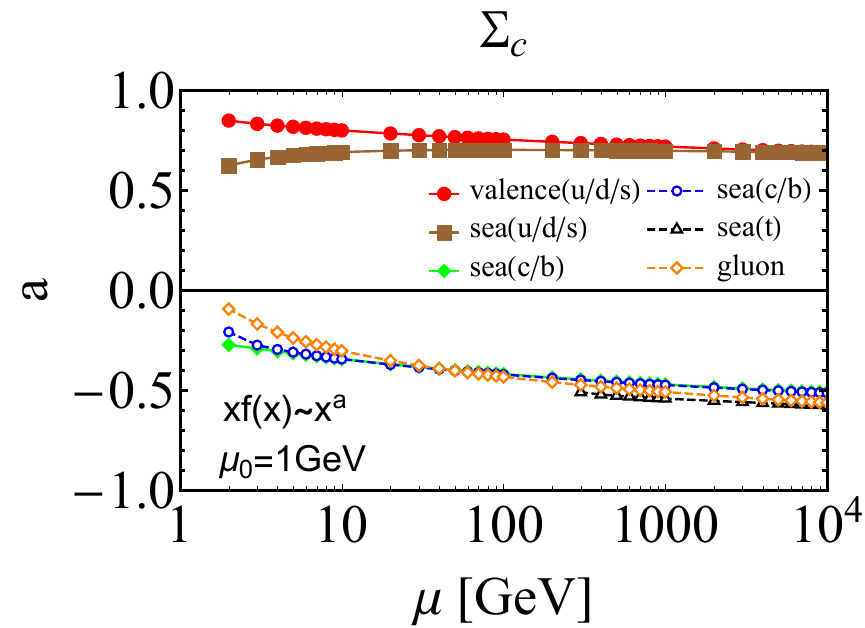}	
	\caption{The exponent $a$ as a function of $\mu$. At a low $x$ ($0.001<x<0.1$), the $x$-PDFs behave as $x f(x)\sim x^a$ for $\Lambda$ (upper-left panel),  $\Sigma$ (upper-right panel, $\Lambda_c$ (lower-left panel and $\Sigma_c$ (lower-right panel).  The red, green, blue, black and orange lines represent valence quark ($u/d$), sea quark ($u/d/s$), sea quark ($c/b$), sea quark ($t$), and gluon, respectively. The dark-yellow lines in the lower pannel represent the valence quark ($c$).}
	\label{fig:low-x}
\end{figure*}
Figure~\ref{figs:lambda-xpdf} shows our results for the valence quark unpolarized PDFs of the $\Lambda$ and its isospin triplet states at the model scale computed using the LFWFs given in Eq.~(\ref{wavefunctions}). The red bands correspond to the results for the light quark ($u$ and/or $d$), whereas the blue bands represent the results for the strange quark. The bands shown in our results arise from the $10\%$ uncertainties in the coupling constant $\alpha_s$. Since $m_s>m_{u(d)}$, the peak of the strange quark distribution in the baryons appears at a higher $x$ compared to the light quark distribution. Therefore, the strange quark carries larger longitudinal momentum than the light quark, reducing the probability of finding a light quark with high $x$ in the baryons. We notice that the valence quark distributions in $\Sigma^+$ and $\Sigma^-$  are nearly identical due to isospin symmetry in the model. However, the magnitude of light quark PDF in these baryons is larger than that of the strange quark PDF. This is due to the fact that there are two up (down) quarks in $\Sigma^+ \,(\Sigma^-)$. Meanwhile, the magnitude of the single light quark distribution is lower compared to that for the strange quark in $\Lambda$ and $\Sigma^0$. 

The valence quark PDFs of the $\Lambda_c\,(\Sigma_c^+,\,\Sigma_c^{++},\,\Sigma_c^0)$ at the initial scale are shown in Fig.~\ref{figs:lambdac-xpdf}. The peaks of
the light quark PDFs again appear at lower $x$, whereas due to the
heavier mass, the charm quark distributions have the peaks
at higher $x$. The PDFs of the $\Lambda_c$ and $\Sigma_c^{++}$ are identical to the distributions of $\Sigma_c^+$ and $\Sigma_c^0$, respectively. Note that $\Sigma_c^{++}\,(\Sigma_c^0)$ has two light quarks of the same flavor, which effectively provides the light quark PDF in $\Sigma_c^{++}\,(\Sigma_c^0)$ twice that of the light quark PDF in $\Lambda_c\,(\Sigma_c^+)$. Figures~\ref{figs:lambda-xpdf}, \ref{figs:lambdac-xpdf} suggest that our model maintains the isospin symmetry, which is also observed from FFs in Figs.~\ref{fig:lambda-ff1}, \ref{fig:lambda-ff2}, \ref{fig:comparison_ff-Sigma}.

We demonstrate the scale evolution of the PDFs of $\Lambda(\Sigma^0,\Sigma^+,\Sigma^-)$  and $\Lambda_c\,(\Sigma_c^+,\,\Sigma_c^{++},\,\Sigma_c^0)$ from the initial scales
to $10$ GeV$^2$ in Figs.~\ref{fig:lambda-10GeV2}
and \ref{fig:lambdac-10GeV2}, respectively. We observe that for $\Lambda$, $\Lambda_c$, and their isospin states, the valence quark distributions increase slowly at lower-$x$ while they decrease at higher-$x$, with the scale evolution. The gluon and the sea
quark PDFs at low-$x$ increase much faster than the valence
quark PDFs. Effectively, in the low-$x$ region the distributions are mainly dominated by the gluon PDFs, while at
large-$x$ the valence quarks dominate the distributions. 
We notice that  the qualitative behavior of the gluon and the sea quarks PDFs obtained by the evolution in both
$\Lambda$, $\Lambda_c$, and their isospin states are very similar.
Since the masses of the charm
and light quarks in $\Lambda_c\,(\Sigma_c^+,\,\Sigma_c^{++},\,\Sigma_c^0)$ are very different, the peaks of their
distributions appear at different $x$.
 The valence quark PDFs in these states exhibit distinctly different behavior compared to that in $\Lambda\,(\Sigma^0,\Sigma^+,\Sigma^-)$, where the valence quark (light and strange) PDFs are close to each other after QCD evolution.
  
The first moments of the corresponding PDFs of $\Lambda$, $\Lambda_c$, and their isospin states as functions
of scale $\mu$ are shown in Fig.~\ref{fig:lambda-x1} and \ref{fig:lambdac-x1}. We notice that the uncertainty bands in Fig.~\ref{fig:lambdac-x1} are insignificant. This is due to the small $\alpha_s$ for $\Lambda_c$ ($\Sigma_c$) and thus, our adopted $10\%$ uncertainty in $\alpha_s$ is also small. We find that with increasing scale $\mu$, as for the other cases described above, momenta carried by the valence quarks decrease, and the contributions of the sea quarks and gluon to the total momentum increase. For the light baryons, the momentum carried by the valence quarks falls faster with increasing scale than that for the baryons having a charm quark. However, the qualitative behaviors of the total moments of valence quarks, sea quarks, and gluons in all states are alike.

Note that the $x$-PDFs at low-$x$ behave like $x^a$, where $a>0$ for the valence quarks, whereas for the gluon and the sea quarks $a>-1$~\cite{Lan:2019img}. The value of the exponent $a$ decreases with increasing scale $\mu$. When $\mu\rightarrow \infty$, $a\rightarrow 0$ for the valence quarks, and for the gluon and the sea quarks, $a\rightarrow -1$. This phenomenon does not depend on the PDFs at the model scale. To illustrate the low-$x$ behavior of the sea quarks and the gluon PDFs with increasing scales, we consider $x$-PDFs of the baryons at low-$x$, $xf(x)\sim x^a$ and present the behavior of $a$ as a function of $\mu$ in Fig.~\ref{fig:low-x}. We observe that $a$ falls steadily and faster for the gluon than that for the sea quarks with increasing $\mu$. This feature again indicates that at low-$x$ the gluon dominates the distribution as the scale increases.

\section{Conclusions}\label{sec:conclusion}
Using a recently proposed light-front model for the baryon based on a Hamiltonian formalism, we presented a comprehensive study of the masses, electromagnetic properties, and parton distribution functions (PDFs) of $\Lambda$, $\Lambda_c$, and their isospin triplet states. The effective Hamiltonian incorporates confinement in both the
transverse and the longitudinal directions, and one-gluon
exchange interaction for the constituent valence quarks suitable for low-resolution properties. We obtained the masses of baryons and the corresponding light-front wave functions (LFWFs) as the eigenvalues and the eigenvectors of this Hamiltonian, respectively by solving its mass eigenvalue equation using basis light-front quantization (BLFQ) as a relativistic three-quark problem. We then employed the LFWFs to investigate the baryon electromagnetic properties and PDFs. We evaluated
the electromagnetic form factors for the baryons and
their flavor decompositions, the magnetic moments, the electric radii
and the magnetic radii of the baryons. We compared our BLFQ results with other theoretical calculations~\cite{Lin:2008mr,VanCauteren:2003hn,Kubis:2000aa,Puglia:2000jy,Julia-Diaz:2004yqv,Julia-Diaz:2004yqv,Faessler:2006ft,Sharma:2010vv,Barik:1983ics,Bernotas:2012nz,Zhu:1997as,Kumar:2005ei,Patel:2007gx,Yang:2018uoj,Wang:2018gpl,Wang:2018gpl}, and experimental data~\cite{GoughEschrich:1998tx,ParticleDataGroup:2020ssz} and found reasonable agreement with available measured data, lattice QCD simulation~\cite{Lin:2008mr}, constituent quark model (CQM)~\cite{VanCauteren:2003hn}, and heavy baryon chiral perturbation theory~\cite{Kubis:2000aa}.

We also computed the unpolarized PDFs of these baryons at a low-resolution scale using our LFWFs. The PDFs at higher scales relevant to experiment and to global QCD analyses have been evaluated based on the NNLO DGLAP equations. The QCD evolution of the PDFs,
which gives us the knowledge of the gluon and the
sea quark distributions, has also been explored. We
observed that although the valence quark dominates at the
large $x>0.1$ domain, at the small-$x$ region the distributions
are mainly controlled by the gluon distribution. The
momenta carried by the gluon and sea quark increase with
increasing scale $\mu$. We also noticed that there is an impression  of universality of the gluon PDFs from different baryon states.
Overall, the QCD evolution of the valence quark PDFs provides
predictions for a wealth of information on the gluon and the
sea quarks arising from higher Fock components. For further
improvement, future developments should focus on the
inclusion of higher Fock sectors directly in the Hamiltonian eigenvalue problem in order to  explicitly incorporate gluon and sea degrees of freedom at appropriate
initial scales. Our work provides predictions for PDFs of the baryons having one strange or charm quark, $\Lambda\,(\Sigma^0,\,\Sigma^+,\,\Sigma^-)$, and $\Lambda_c\,(\Sigma_c^+,\,\Sigma_c^{++},\,\Sigma_c^0)$ from future experiments as well as a guidance for the theoretical
studies of the PDFs with higher Fock sectors.

Since our LFWFs incorporate all three active quarks’ spin, flavor, and three-dimensional spatial information on the same footing, the effective LFWFs can be employed to investigate other parton distributions, such as the helicity and tranversity PDFs, the generalized parton distributions, the transverse momentum dependent parton distributions, the Wigner distributions, and double parton distribution functions as well as mechanical properties of the baryons. The presented results affirm
the utility of our model and motivate application of
analogous effective Hamiltonians to other heavy baryons.
\\

\begin{acknowledgments}
We thank Jiangshan Lan and Zhi Hu for many useful discussions. C. M. is supported by new faculty start up funding by the Institute of Modern Physics, Chinese Academy of Sciences, Grant No. E129952YR0. 
C. M. also thanks the Chinese Academy of Sciences Presidents International Fellowship Initiative for the support via Grants No. 2021PM0023. X. L. is supported by the China National Funds for Distinguished Young Scientists under Grant No. 11825503, National Key Research and Development Program of China under Contract No. 2020YFA0406400, the 111 Project under Grant No. B20063, and the National Natural Science Foundation of China under Grant No. 12047501. X. Z. is supported by new faculty startup funding by the Institute of Modern Physics, Chinese Academy of Sciences, by Key Research Program of Frontier Sciences, Chinese Academy of Sciences, Grant No. ZDB-SLY-7020, by the Natural Science Foundation of Gansu Province, China, Grant No. 20JR10RA067, by the Foundation for Key Talents of Gansu Province, by the Central Funds Guiding the Local Science and Technology Development of Gansu Province and by the Strategic Priority Research Program of the Chinese Academy of Sciences, Grant No. XDB34000000. J. P. V. is supported by the Department of Energy under Grants No. DE-FG02-87ER40371, and No. DE-SC0018223 (SciDAC4/NUCLEI). This research used resources of the National Energy Research Scientific Computing Center (NERSC), a U.S. Department of Energy Office of Science User Facility operated under Contract No. DE-AC02-05CH11231.
 A portion of the computational resources were also provided by Gansu Computing Center.
\end{acknowledgments}

\begin{appendices}
	
\section{APPENDIX: THE SPIN-FLAVOR STRUCTURES OF $\Lambda$($\Lambda_c$) and $\Sigma$($\Sigma_c$) BARYONS}
\begin{align*}
|\Lambda,\uparrow\rangle_{\text{F $\otimes$ S}}=&\frac{1}{\sqrt{2}}\Big[\frac{1}{2}(sud+usd-sdu-dsu)\nonumber\\
&\otimes\frac{1}{\sqrt{6}} (\uparrow\downarrow\uparrow+\downarrow\uparrow\uparrow-2\uparrow\uparrow\downarrow) \nonumber\\
+&\frac{1}{\sqrt{12}}(dsu-sdu+sud-usd+2uds-2dus)\nonumber\\
&\otimes\frac{1}{\sqrt{2}} (\uparrow\downarrow\uparrow-\downarrow\uparrow\uparrow)\Big],\nonumber\\
|\Sigma^0,\uparrow\rangle_{\text{F $\otimes$ S}}=&\frac{1}{\sqrt{2}}\Big[ \frac{1}{2\sqrt{3}}(sdu+sud+uds+dsu-2uds-2dus)\nonumber\\
&\otimes\frac{1}{\sqrt{6}} (\uparrow\downarrow\uparrow+\downarrow\uparrow\uparrow-2\uparrow\uparrow\downarrow) \nonumber\\
+&\frac{1}{2}(sud-usd-dsu+sdu)\nonumber\\
&\otimes\frac{1}{\sqrt{2}} (\uparrow\downarrow\uparrow-\downarrow\uparrow\uparrow)\Big].\nonumber\\
|\Lambda_c,\uparrow\rangle_{\text{F $\otimes$ S}}=&\frac{1}{\sqrt{2}}\Big[\frac{1}{2}(cud+ucd-cdu-dcu)\nonumber\\
&\otimes\frac{1}{\sqrt{6}} (\uparrow\downarrow\uparrow+\downarrow\uparrow\uparrow-2\uparrow\uparrow\downarrow) \nonumber\\
+&\frac{1}{\sqrt{12}}(dcu-cdu+cud-ucd+2udc-2duc)\nonumber\\
&\otimes\frac{1}{\sqrt{2}} (\uparrow\downarrow\uparrow-\downarrow\uparrow\uparrow)\Big],\nonumber\\
|\Sigma_c^+,\uparrow\rangle_{\text{F $\otimes$ S}}=&\frac{1}{\sqrt{2}}\Big[ \frac{1}{2\sqrt{3}}(cdu+cud+udc+dcu-2udc-2duc)\nonumber\\
&\otimes\frac{1}{\sqrt{6}} (\uparrow\downarrow\uparrow+\downarrow\uparrow\uparrow-2\uparrow\uparrow\downarrow) \nonumber\\
+&\frac{1}{2}(cud-ucd-dcu+cdu)\nonumber\\
&\otimes\frac{1}{\sqrt{2}} (\uparrow\downarrow\uparrow-\downarrow\uparrow\uparrow)\Big].
\label{spin_flavour_wave_function}
\end{align*}
\end{appendices}


\bibliography{references.bib}

\end{document}